\documentclass[10pt, draftcls, onecolumn, journal]{IEEEtran}
\usepackage{cite, graphicx,amssymb,color, kbordermatrix, slashbox, pict2e, multirow, rotating}
\usepackage{graphicx,amssymb}
\usepackage[tbtags]{amsmath}
\usepackage{times,amsmath}
\usepackage{subfig}
\usepackage{flushend}
\usepackage{amsmath}
\usepackage{epsfig}
\usepackage{amssymb}
\usepackage{hyperref}
\usepackage{color}
\usepackage{psfrag}
\usepackage{caption}
\usepackage{subfig}
\usepackage[usenames,dvipsnames]{xcolor}
\usepackage{textcomp}

\usepackage{tikz}
\usetikzlibrary{matrix,decorations.pathreplacing}

\begin{document}
\vspace{-10mm}
\title{Efficient Strategies for Single/Multi-Target Jamming on MIMO Gaussian Channels}
\vspace{-2mm}
\author{{Jie Gao,
Sergiy A. Vorobyov,
and Hai Jiang 
}
\vspace{-2mm}
}

\maketitle

\begin{abstract}
The problem of jamming on multiple-input multiple-output (MIMO) Gaussian channels is investigated in this paper. In the case of a single target legitimate signal, we show that the existing result based on the simplification of the system model by neglecting the jamming channel leads to losing important insights regarding the effect of jamming power and jamming channel on the jamming strategy. We find a closed-form optimal solution for the problem under a positive semi-definite (PSD) condition without considering simplifications in the model. If the condition is not satisfied and the optimal solution may not exist in closed-form, we find the optimal solution using a numerical method and also propose a suboptimal solution in closed-form as a close approximation of the optimal solution. Then, the possibility of extending the results to solve the problem of multi-target jamming is investigated for four scenarios, i.e., multiple access channel, broadcasting channel, multiple transceiver pairs with orthogonal transmissions, and multiple transceiver pairs with interference, respectively. It is shown that the proposed numerical method can be extended to all scenarios while the proposed closed-form solutions for jamming may be applied in the scenarios of the multiple access channel and multiple transceiver pairs with orthogonal transmissions. Simulation results verify the effectiveness of the proposed solutions.
\end{abstract}

\vspace{-0.1cm}

\IEEEpeerreviewmaketitle

\section{Introduction}
Security is a major concern in wireless communications \cite{Intr1}-\cite{Intr4}. Due to the rapid development of wireless communications, the security issue rises while wireless communication networks of different scales containing devices for different purposes become more common and popular. Major threats to wireless communications include  passive wiretapping and active jamming \cite{Intr5}. While the passive threat can be addressed by using well-designed security architectures, wireless communications are vulnerable to the active jamming attack \cite{IntJ1}. Jamming aims at degrading the quality of communication or disrupting the information transmission in a communication system by directing energy toward the target receiver in a destructive manner \cite{JamBook1}. A jamming attack is particularly effective because it is easy to launch using low-cost and small-sized devices while causing very significant security threat \cite{Jam3}.  The threat of jamming has been studied in many research works \cite{Jam1.2, Jam1.4, Jam1.8}, and one of the relevant research interests is to investigate the optimal jamming strategy from the perspective of a jammer \cite{Jam3, Jam1, Jam2}. Such perspective helps to reveal the effect of jamming on legitimate communications in the worst case.

When a jammer has multiple antennas, it can maximize the effectiveness of jamming by optimizing its jamming signal. The optimal jamming on multiple-input multiple-output (MIMO) channels is investigated in \cite{JamMIMO1}-\cite{JamMIMO3}. It is shown in \cite{JamMIMO1} that without knowledge of the target signal or its covariance, the jammer can only use basic strategies of allocating power uniformly or maximizing the total power of the interference at the target receiver. In \cite{JamMIMO1.8}, the transmit strategies of a legitimate transmitter and a jammer on a Gaussian MIMO channel are investigated under a game-theoretic modeling with a general utility function. It is assumed that the jammer and the legitimate transmitter have the same level of channel state information (CSI), i.e., both uninformed, both with statistical CSI, or both with exact CSI. The optimal transmitted strategies of the legitimate transmitter and the jammer are represented as solutions to different optimization problems versus different types of CSI. The worst-case jamming on MIMO multiple access and broadcast channels with the covariance of the target signal and all channel information available at the jammer is studied in \cite{JamMIMO2} based on game theory. Some properties of the optimal jamming strategies are characterized through the analysis of the Nash equilibrium of the game. The  necessary condition for optimal jamming on MIMO channels with arbitrary inputs when the covariance of the target signal and all channel information are available at the jammer is derived in \cite{JamMIMO3}. For the case of Gaussian target signal, the solution of optimal jamming is given in closed-form. However, it is derived without considering the jamming channel.  As a result, the system model is oversimplified by implicitly assuming that the received jamming signal at the target receiver is exactly the same as the transmitted jamming signal at the jammer.

With the objective of providing a general solution without oversimplifications of the system model, this work addresses the problem of optimal jamming on MIMO Gaussian channels. First, the problem of jamming a single target communication between a legitimate transceiver pair will be investigated. Then, we show that the methods used for obtaining the solution of the single-target jamming problem can be extended to solve multi-target jamming problem. The main contributions of this work are as follows.

First, for the general case of jamming a target communication on a MIMO Gaussian channel, we show that the optimal solution may or may not exist in closed-form. It is shown that the existence of the optimal solution in closed-form, given the target signal and the legitimate communication channel, depends on the jammer's power limit and the jamming channel. The optimal solution in closed-form is given under a positive semi-definite (PSD) condition and the solution in \cite{JamMIMO3} is shown to be a special case of our general solution.

Second, we propose a suboptimal solution in closed-form as an alternative strategy for the jammer so that the complexity of finding the solution remains low when there is no closed-form expression for the optimal solution. For finding the optimal solution in this case, a numeric method which is proved to converge to optimality is used. The two alternative solutions provide a choice between low complexity and high accuracy. Simulation demonstrates that the proposed suboptimal solution is in fact very close to the optimal one, and thereby qualifies as a very good approximation of the optimal solution.

Third, we extend the above results by considering multi-target jamming. Four scenarios of multi-target jamming are considered, i.e.,  jamming a multiple access channel, jamming a broadcast channel, jamming multiple transceiver pairs with orthogonal transmissions, and jamming multiple transceiver pairs with interference. It is shown that while the numeric method for finding the optimal solution to the problem of single target jamming can be extended, after proper modifications, to all four scenarios, the methodology in obtaining the closed-form expressions of the optimal/suboptimal solution can be adopted for the scenario of jamming a multiple access channel and possibly the scenario of jamming multiple transceiver pairs with orthogonal transmissions.

The rest of the paper is organized as follows. Section~\ref{s:sysm} gives the system model of this work. The closed-form solution to the problem of jamming a single legitimate communication and the condition for it to exist are investigated in Section~\ref{C5ss:OneCloseF}. When this condition is not met, the optimal numerical solution and a suboptimal solution in closed-form are found in Sections~\ref{s:numer}~and~\ref{s:subsolu}, respectively. The possibility of extending the results to multi-target jamming and the corresponding modifications required are demonstrated in Section~\ref{s:mutlijam}. Section~\ref{s:simula} shows the simulation results which verify the effectiveness of the solutions obtained in previous sections and Section~\ref{s:conclu} concludes the
paper. Section~\ref{s:appen} ``Appendix'' provides proofs for the lemmas and theorems.

\section{System Model}\label{s:sysm}
A legitimate transmitter with $n_\mathrm{t}$ antennas sends a signal $\mathbf{s}$ to a receiver with $n_\mathrm{r}$ antennas. The elements of $\mathbf{s}$ are independent and identically distributed Gaussian with zero mean and covariance $\mathbf{Q}_\mathrm{s}$. A jammer with $n_\mathrm{z}$ antennas attempts to jam the legitimate communication by transmitting a jamming signal $\mathbf{z}$ to the receiver. Denote the legitimate channel from the legitimate transmitter to the receiver as $\mathbf{H}_\mathrm{r}$ (of size $n_\mathrm{r}\times n_\mathrm{t}$) and the jamming channel from the jammer to the receiver as $\mathbf{H}_\mathrm{z}$ (of size $n_\mathrm{r}\times n_\mathrm{z}$). In the presence of the jamming signal, the received signal at the legitimate receiver is expressed as
\begin{equation}\label{e:NoCJSig}
\mathbf{y}=\mathbf{H}_\mathrm{r} \mathbf{s} + \mathbf{H}_\mathrm{z}\mathbf{ z} + \mathbf{n}
\end{equation}
where $\mathbf{n}$ is the noise at the legitimate receiver with zero mean and covariance $\sigma^2\mathbf{I}$. Here $\mathbf{I}$ denotes the identity matrix of an appropriate size. Note that given the Gaussian channel and Gaussian target signal, the worst-case form of jamming signal is also Gaussian \cite{Worstnoise}. Denote the covariance of $\mathbf{z}$ as $\mathbf{Q}_\mathrm{z}$. Then the information rate of the legitimate communication in the presence of the jamming is expressed at
\begin{equation}\label{e:NoCJRate}
R^\mathrm{J}=\log|\mathbf{I}+\mathbf{H}_\mathrm{r}\mathbf{Q}_\mathrm{s}\mathbf{H}_\mathrm{r}^\mathrm{H}(\mathbf{H}_\mathrm{z}
\mathbf{Q}_\mathrm{z}\mathbf{H}_\mathrm{z}^\mathrm{H} +\sigma^2\mathbf{I})^{-1}|
\end{equation}
where $|\cdot|$ and $(\cdot)^\mathrm{H}$ denote the determinant and the Hermitian transpose, respectively. The jammer aims at decreasing the above rate as much as possible given its power limit $P_\mathrm{z}$. The jammer is assumed to have the knowledge of $\mathbf{H}_\mathrm{r}$, $\mathbf{H}_\mathrm{z}$, and $\mathbf{Q}_\mathrm{s}$ but not the exact $\mathbf{s}$. As a result, it is not able to perform correlated jamming \cite{CorreJamingFading}. However, the jammer can use the available knowledge to find the optimal $\mathbf{Q}_\mathrm{z}$ such that the rate \eqref{e:NoCJRate} is minimized. This problem is studied in details in the following section.

\section{Optimal jamming in closed-form under PSD condition}\label{C5ss:OneCloseF}
Given the system model, the optimal jamming strategy can be found by solving the following problem\footnote{The PSD constraint $\mathbf{Q}_\mathrm{z}\succeq0$ is assumed by default and it is omitted for brevity throughout this section.}
\begin{subequations}\label{e:JOne} % Jam one terminal
\begin{align}
& \mathop{\mathbf{min}}\limits_{\mathbf{Q}_\mathrm{z}}  \quad R^\mathrm{J}  \label{e:JOneobj} \\
& \mathbf{\;s.t.}  \quad\quad  \text{Tr}\{\mathbf{Q}_\mathrm{z}\}\leq P_\mathrm{z} \label{e:JOnecon}
\end{align}
\end{subequations}
where $\text{Tr}\{\cdot\}$ denotes the trace. With only one pair of transceiver, the above problem is a basic jamming problem on a MIMO channel.

Denote the singular value decomposition (SVD) of $\mathbf{H}_\mathrm{z}$ as $\mathbf{H}_\mathrm{z}=\mathbf{U}_\mathrm{z}\mathbf{\Omega}_\mathrm{z}\mathbf{V}_\mathrm{z}^\mathrm{H}$. The matrices $\mathbf{U}_\mathrm{z}$, $\mathbf{\Omega}_\mathrm{z}$, and $\mathbf{V}_\mathrm{z}$ are of sizes $n_\mathrm{r}\times n_\mathrm{r}$, $n_\mathrm{r}\times n_\mathrm{z}$, and $n_\mathrm{z}\times n_\mathrm{z}$, respectively. Define $\mathbf{B}\triangleq\mathbf{U}_\mathrm{z}^\mathrm{H}
\mathbf{H}_\mathrm{r}\mathbf{Q}_\mathrm{s}\mathbf{H}_\mathrm{r}^\mathrm{H}\mathbf{U}_\mathrm{z}$. Note that $\mathbf{B}$ has the same rank as $\mathbf{H}_\mathrm{r}\mathbf{Q}_\mathrm{s}\mathbf{H}_\mathrm{r}^\mathrm{H}$. Using the definition of $\mathbf{B}$ and the SVD of $\mathbf{H}_\mathrm{z}$, the objective function in \eqref{e:NoCJRate} can be rewritten as
\begin{equation}\label{e:JOneobjEqv1t} 
R^\mathrm{J}=\log{|\mathbf{I}+\mathbf{B}(\mathbf{\Omega}_\mathrm{z}
\hat{\mathbf{Q}}_\mathrm{z}\mathbf{\Omega}_\mathrm{z}^\mathrm{H}+\sigma^2\mathbf{I})^{-1}|}
\end{equation}
where
\begin{equation}\label{e:Qhatdeft}
\hat{\mathbf{Q}}_\mathrm{z}\triangleq\mathbf{V}_\mathrm{z}^\mathrm{H}\mathbf{Q}_\mathrm{z}\mathbf{V}_\mathrm{z}.
\end{equation}

In order to solve the optimization problem \eqref{e:JOne}, we start from introducing the following two lemmas.

\emph{Lemma~1}: Given a constant Hermitian matrix $\mathbf{A}$ with $\mathbf{A}\succ0$, the optimization problem over positive definite (PD) matrix $\mathbf{X}$
\begin{subequations} \label{e:BJ}%basic jamming problem
\begin{align}
& \mathop{\mathbf{min}}\limits_{\mathbf{X}}  \quad \log{|\mathbf{I}+\mathbf{A}\mathbf{X}^{-1}|} \label{e:BJobj}\\
& \mathbf{\;s.t.}  \quad\quad  \text{Tr}\{\mathbf{X}\}\leq 1 \label{e:BJcons1}\\
& \qquad\qquad \mathbf{X} \succeq 0 \label{e:BJcons2}
\end{align}
\end{subequations}
has the following closed-form solution
\begin{equation}\label{e:BJsolu}
\mathbf{X}=\mathbf{U}_{\rm A}\sqrt{\frac{\mathbf{\Lambda}_{\rm A}}{\lambda}+\frac{\mathbf{\Lambda}_{\rm A}^2}{4}}\mathbf{U}_{\rm A}^{\rm H}-\frac{\mathbf{A}}{2}
\end{equation}
where $\mathbf{U}_{\rm A}$ and $\mathbf{\Lambda}_{\rm A}$ are the eigenvector and eigenvalue matrices, respectively, obtained from the eigenvalue decomposition (EVD) $\mathbf{A}=\mathbf{U}_{\rm A}\mathbf{\Lambda}_{\rm A}\mathbf{U}_{\rm A}^{\rm H}$, and $\lambda$ is chosen so that the power constraint \eqref{e:BJcons1} is satisfied with equality.

\textbf{Proof}: See Subsection~\ref{App:C5prfl1} in Appendix.

Lemma~1 gives the closed-form solution to problem \eqref{e:BJ}, which is similar to but simpler than problem \eqref{e:JOne}. However, it can be seen that the obtained solution cannot be straightforwardly extended to obtain the solution to problem \eqref{e:JOne}. Indeed, the two terms, i.e., $\mathbf{\Omega}_\mathrm{z}$ and $\mathbf{\Omega}_\mathrm{z}^\mathrm{H}$ multiplied to $\hat{\mathbf{Q}}_\mathrm{z}$ in \eqref{e:JOneobjEqv1t} leads to a more complicated solution, especially considering that the matrix $\mathbf{\Omega}_\mathrm{z}$ can be rank deficient. Nevertheless, as will be shown later, the solution \eqref{e:BJsolu} to problem \eqref{e:BJ} will help in deriving the solution to problem \eqref{e:JOne}.

Denote the rank of $\mathbf{H}_\mathrm{z}$ as $r_\mathrm{z}$ and assume without loss of generality that the first $r_\mathrm{z}$ elements on the main diagonal of $\mathbf{\Omega}_\mathrm{z}$ are non-zero. Whether or not $\mathbf{B}$ is PD, i.e., has the rank of $n_\mathrm{z}$, has an impact on the optimal form of $\hat{\mathbf{Q}}_\mathrm{z}$ in \eqref{e:JOneobjEqv1t}. Therefore, the following lemma regarding $\mathbf{B}$ is in order.

\emph{Lemma~2}: If we denote $\mathbf{B}$ using blocks such that
\begin{equation}\label{e:Bblocks}
\mathbf{B}=
\kbordermatrix{
    ~   & r_\mathrm{z} & n_\mathrm{z}-r_\mathrm{z} \cr
    r_\mathrm{z}   & {\mathbf{B}_{11}}   &  {\mathbf{B}_{12}}  \cr
    n_\mathrm{z}-r_\mathrm{z} &  {\mathbf{B}_{21}}  &  {\mathbf{B}_{22}}   \cr}
\end{equation}
and define
\begin{equation}\label{e:tildeB}
\mathbf{\tilde{B}}\triangleq{\mathbf{B}_{11}}- {\mathbf{B}_{12}}(\sigma^2\mathbf{I}+{\mathbf{B}_{22}})^{-1}{\mathbf{B}_{21}},
\end{equation}
then $\mathbf{\tilde{B}}$ is PD if $\mathbf{B}$ is PD.

\textbf{Proof}: See Subsection~\ref{App:C5prfl2} in Appendix.

Before solving the optimization problem \eqref{e:JOne} based on the above two lemmas, it is essential to express the objective function of problem \eqref{e:JOne} in a different form so as to reveal the optimal structure of $\mathbf{Q}_\mathrm{z}$. Denote the diagonal matrix $\mathbf{\Omega}_\mathrm{z}$ using blocks as
\begin{equation}\label{e:Omgz}
\mathbf{\Omega}_\mathrm{z}\triangleq
    \kbordermatrix{
    ~   & r_\mathrm{z} & n_\mathrm{z}-r_\mathrm{z} \cr
    r_\mathrm{z}   & {\mathbf{\Omega}_\mathrm{z}^{+}} & \mathbf{0} \cr
    n_\mathrm{r}-r_\mathrm{z} &  \mathbf{0} &\mathbf{0}  \cr}
\end{equation}
where $\mathbf{\Omega}_\mathrm{z}^{+}$ is an $r_\mathrm{z} \times r_\mathrm{z}$ diagonal matrix made of the positive diagonal elements of $\mathbf{\Omega}_\mathrm{z}$, and $\mathbf{0}$ denotes an all-zero matrix of appropriate size. It can be seen that the allocation of jamming power should be limited to at most $r_\mathrm{z}$ dimensions corresponding to the $r_\mathrm{z}$ non-zero eigenvalues of $\mathbf{\Omega}_\mathrm{z}$. Indeed, allocating jamming power anywhere else has no effect on the received signal and only leads to jamming power waste. As a result, $\mathbf{\hat{Q}}_\mathrm{z}$ should adopt the following form
\begin{eqnarray}\label{e:Qhatzb}
\vspace{-1mm}
\mathbf{\hat{Q}}_\mathrm{z}=
\kbordermatrix{
    ~   & r_\mathrm{z} & n_\mathrm{z}-r_\mathrm{z} \cr
    r_\mathrm{z}   & \mathbf{Q}^{\prime}_\mathrm{z}   &  \mathbf{\Gamma}_\mathrm{z}  \cr
    n_\mathrm{z}-r_\mathrm{z} &  \mathbf{\Gamma}_\mathrm{z}^\mathrm{H}  &  \mathbf{0}   \cr}
    \vspace{-1mm}
\end{eqnarray}
where $\mathbf{Q}^{\prime}_\mathrm{z}$ and $\mathbf{\Gamma}_\mathrm{z}$ are to be determined. It can be shown that the specific matrix $\mathbf{\Gamma}_\mathrm{z}$ does not affect the rate of $R^\mathrm{J}$ in \eqref{e:JOneobjEqv1t}. Therefore, $\mathbf{\Gamma}_\mathrm{}z$ is set to be $\mathbf{0}$ for simplicity and consequently
\begin{eqnarray}\label{e:Qhatblockt}
\vspace{-1mm}
\mathbf{\hat{Q}}_\mathrm{z}=
\kbordermatrix{
    ~   & r_\mathrm{z} & n_\mathrm{z}-r_\mathrm{z} \cr
    r_\mathrm{z}   & \mathbf{Q}^{\prime}_\mathrm{z}   &  \mathbf{0}  \cr
    n_\mathrm{z}-r_\mathrm{z} &  \mathbf{0}  &  \mathbf{0}   \cr}.
    \vspace{-1mm}
\end{eqnarray}

Let us define a new eigen channel $\mathbf{\tilde{\Omega}}_\mathrm{z}$ as
\begin{equation}\label{e:OmgzTt}
\mathbf{\tilde{\Omega}}_\mathrm{z}\triangleq
    \kbordermatrix{
    ~   & r_\mathrm{z} & n_\mathrm{r}-r_\mathrm{z} \cr
    r_\mathrm{z}   & {\mathbf{\Omega}_\mathrm{z}^{+}} & \mathbf{0} \cr
    n_\mathrm{r}-r_\mathrm{z} &  \mathbf{0} &\mathbf{I}  \cr}.
\end{equation}
The equivalent channel matrix $\mathbf{\tilde{\Omega}}_\mathrm{z}$ has size $n_\mathrm{r} \times n_\mathrm{r}$, which is larger than the size of $\mathbf{\Omega}_\mathrm{z}$ if $n_\mathrm{r}>n_\mathrm{z}$, smaller than the size of $\mathbf{\Omega}_\mathrm{z}$ if $n_\mathrm{r}<n_\mathrm{z}$ and has the same size as $\mathbf{\Omega}_\mathrm{z}$ if $n_\mathrm{r}=n_\mathrm{z}$. Also define the following new jamming covariance matrix $\mathbf{\tilde{Q}}_\mathrm{z}$ as
\begin{eqnarray}\label{e:Qtildet}
\vspace{-1mm}
\mathbf{\tilde{Q}}_\mathrm{z}\triangleq
\kbordermatrix{
    ~   & r_\mathrm{z} & n_\mathrm{r}-r_\mathrm{z} \cr
    r_\mathrm{z}   & \mathbf{Q}^{\prime}_\mathrm{z}   &  \mathbf{0}  \cr
    n_\mathrm{r}-r_\mathrm{z} &  \mathbf{0}  &  \mathbf{0}   \cr}
    \vspace{-1mm}
\end{eqnarray}
where $\mathbf{Q}^{\prime}_\mathrm{z}$ is the same as in \eqref{e:Qhatzb}.

With the above definitions of $\mathbf{\tilde{\Omega}}_\mathrm{z}$ and $\mathbf{\tilde{Q}}_\mathrm{z}$, it can be seen that $\mathbf{\Omega}_\mathrm{z}
\hat{\mathbf{Q}}_\mathrm{z}\mathbf{\Omega}_\mathrm{z}^\mathrm{H}$ in \eqref{e:JOneobjEqv1t} is equal to $\mathbf{\tilde{\Omega}}_\mathrm{z}
\mathbf{\tilde{Q}}_\mathrm{z}\mathbf{\tilde{\Omega}}_\mathrm{z}^\mathrm{H}$. As a result, the rate in \eqref{e:JOneobjEqv1t} can be equivalently rewritten as
\begin{equation}\label{e:JoneobjEqvt2}
\hspace{-0.7cm}R^{\rm J}=\log{|\mathbf{I}+\mathbf{B}(\mathbf{\tilde{\Omega}}_\mathrm{z}
\mathbf{\tilde{Q}}_\mathrm{z}\mathbf{\tilde{\Omega}}_\mathrm{z}^\mathrm{H}+\sigma^2\mathbf{I})^{-1}|}.
\end{equation}
Therefore, we consider $\mathbf{\tilde{\Omega}}_\mathrm{z}$ and $\mathbf{\tilde{Q}}_\mathrm{z}$ as the equivalent channel matrix and the equivalent jamming covariance matrix to $\mathbf{\Omega}_\mathrm{z}$ and $\hat{\mathbf{Q}}_\mathrm{z}$, respectively. The advantage of solving the optimization problem \eqref{e:JOne} using the above equivalent form of the objective function is that $\mathbf{\tilde{\Omega}}_\mathrm{z}$ and $\mathbf{\tilde{\Omega}}_\mathrm{z}^\mathrm{H}$ in \eqref{e:JoneobjEqvt2} are always PD and therefore can be extracted from the inverse term, which simplified the solution finding procedure.

Using the above paragraphs and equations \eqref{e:Qhatdeft} and \eqref{e:Qhatblockt} it can be seen that the optimal form of $\mathbf{Q}_\mathrm{z}$ is
\begin{equation}\label{e:QOptform}
\mathbf{Q}_\mathrm{z}=\mathbf{V}_\mathrm{z}\left[\!\! {\begin{array}{*{20}c}
   {\mathbf{Q}_\mathrm{z}^{\prime}} & {\mathbf{0}}\\
   {\mathbf{0}}&{\mathbf{0}}\\
   \end{array} } \!\!\right]\mathbf{V}_\mathrm{z}^\mathrm{H}.
\end{equation}
where the two diagonal blocks of in the block diagonal matrix have sizes $r_\mathrm{z}\times r_\mathrm{z}$ and $(n_\mathrm{z}-r_\mathrm{z})\times (n_\mathrm{z}-r_\mathrm{z})$, respectively.

Given the above definitions and lemmas, we next solve the problem \eqref{e:JOne} by finding the optimal $\mathbf{Q}_\mathrm{z}^{\prime}$ in \eqref{e:QOptform}. First, we consider the case that $\mathbf{H}_\mathrm{r}\mathbf{Q}_\mathrm{s}\mathbf{H}_\mathrm{r}^\mathrm{H}$ in \eqref{e:NoCJRate} is PD. Then, we will extend the solution to the more general case that $\mathbf{H}_\mathrm{r}\mathbf{Q}_\mathrm{s}\mathbf{H}_\mathrm{r}^\mathrm{H}$ in \eqref{e:NoCJRate} is PSD but not necessarily PD.

\textbf{Theorem~1}: When $\mathbf{H}_\mathrm{r}\mathbf{Q}_\mathrm{s}\mathbf{H}_\mathrm{r}^\mathrm{H}$ is positive definite, the problem \eqref{e:JOne} has the following closed-form optimal solution
\begin{equation}\label{e:QOptS1}
\mathbf{Q}_\mathrm{z}^{\prime}=\mathbf{U}_{\mathbf{\tilde{A}}}\sqrt{\frac{1}{\lambda}\mathbf{\Lambda}_{\mathbf{\tilde{A}}}\!+\!
\frac{1}{4}\mathbf{\Lambda}_{\mathbf{\tilde{A}}}^2}\mathbf{U}_{\mathbf{\tilde{A}}}^{\rm H}-{\mathbf{\Omega}_\mathrm{z}^{+}}^{\!-1}\bigg(\frac{1}{2}\mathbf{\tilde{B}}\!+\! \sigma^2\mathbf{I}\bigg)
{\mathbf{\Omega}_\mathrm{z}^{+}}^{\!-\mathrm{H}}
\end{equation}
under the condition that the above matrix $\mathbf{Q}_\mathrm{z}^{\prime}$ is PSD,
where $\mathbf{\tilde{B}}$ is given by \eqref{e:tildeB}, $\mathbf{U}_{\mathbf{\tilde{A}}}$ and $\mathbf{\Lambda}_{\mathbf{\tilde{A}}}$ are obtained from the EVD $\mathbf{\tilde{A}}=\mathbf{U}_{\mathbf{\tilde{A}}}\mathbf{\Lambda}_{\mathbf{\tilde{A}}}\mathbf{U}_{\mathbf{\tilde{A}}}^\mathrm{H}$
with
\begin{equation}\label{e:tildeA}
\mathbf{\tilde{A}}\triangleq{\mathbf{\Omega}_\mathrm{z}^{+}}^{\!-1}\mathbf{\tilde{B}}{\mathbf{\Omega}_\mathrm{z}^{+}}^{\!-\mathrm{H}},
\end{equation}
and $\lambda$ is chosen such that the jammer's power constraint \eqref{e:JOnecon} is satisfied with equality.

\textbf{Proof}: Please see Section~\ref{App:C5prfT1} in Appendix.

As mentioned in Section~I, a special case of the problem \eqref{e:JOne} that assumes the jamming channel $\mathbf{H}_\mathrm{z}$ to be the identity matrix $\mathbf{I}$ is investigated in \cite{JamMIMO3}. Consequently, $\mathbf{U}_\mathrm{z}$, $\mathbf{\Omega}_\mathrm{z}$, and $\mathbf{V}_\mathrm{z}^\mathrm{H}$ are all equal to $\mathbf{I}$. Therefore, $\mathbf{\tilde{A}}$ and $\mathbf{\Omega}_\mathrm{z}^{+}$ simplify to $\mathbf{\tilde{B}}$ and $\mathbf{I}$, respectively. Moreover, the above simplification in \cite{JamMIMO3} leads to the result that $r_\mathrm{z}=n_\mathrm{z}$, which further simplifies the case so that $\mathbf{\tilde{B}}= \mathbf{B}$ and $\mathbf{Q}_\mathrm{z}=\mathbf{Q}_\mathrm{z}^{\prime}$. Then, \eqref{e:QOptS1} becomes the following simplified solution
\begin{eqnarray}\label{e:QOptS1Special}
\mathbf{Q}_\mathrm{z}^{\prime}=\mathbf{U}_{\mathbf{B}}\bigg(\sqrt{\frac{1}{\lambda}\mathbf{\Lambda}_{\mathbf{B}}+
\frac{1}{4}\mathbf{\Lambda}_{\mathbf{B}}^2}-\frac{1}{2}\mathbf{\Lambda}_\mathbf{B}-\sigma^2\mathbf{I}\bigg)\mathbf{U}_{\mathbf{B}}^{\rm H}
\end{eqnarray}
where $\mathbf{U}_{\mathbf{B}}$ and $\mathbf{\Lambda}_\mathbf{B}$ are obtained from the EVD $\mathbf{B}=\mathbf{U}_{\mathbf{B}}\mathbf{\Lambda}_\mathbf{B}\mathbf{U}_{\mathbf{B}}^\mathrm{H}$. An equivalent scalar form of the above solution is given in \cite{JamMIMO3} for the above oversimplified case of the problem. By forcing the negative elements (if any) of $\sqrt{\mathbf{\Lambda}_{\mathbf{B}}/\lambda+ \mathbf{\Lambda}_{\mathbf{B}}^2/4}-\mathbf{\Lambda}_\mathbf{B}/{2}- \sigma^2\mathbf{I}$ to be zero and adjusting $\lambda$ to satisfy the power constraint, the solution given in \eqref{e:QOptS1Special} can always be made PSD.

The solution of $\mathbf{Q}^{\prime}_\mathrm{z}$ given by \eqref{e:QOptS1} is not necessarily PSD for the case considered in Theorem~1. It can be indefinite when the jammer's power limit $P_\mathrm{z}$ is sufficiently small. It can be seen that $1/\lambda$ decreases when the jammer's power limit becomes smaller. As a result, $\mathbf{Q}_\mathrm{z}^{\prime}$ in \eqref{e:QOptS1} has a larger chance to be indefinite and thereby invalid as a solution of a covariance matrix. For a given power limit $P_\mathrm{z}$, whether or not $\mathbf{Q}^{\prime}_\mathrm{z}$ in \eqref{e:QOptS1} is PSD depends on the elements of $\mathbf{\Omega}_\mathrm{z}^{+}$, or equivalently the channel $\mathbf{H}_\mathrm{z}$. It can be shown that, for a small $P_\mathrm{z}$ and a given $\mathbf{\Omega}_\mathrm{z}^{+}$ such that $\mathbf{Q}^{\prime}_\mathrm{z}$ given by \eqref{e:QOptS1} is indefinite, there always exists $\mathbf{\tilde{\Omega}}_\mathrm{z}^{+}$ with the same trace as $\mathbf{\Omega}_\mathrm{z}^{+}$ (i.e., $\text{Tr}\{\mathbf{\tilde{\Omega}}_\mathrm{z}^{+}\} =\text{Tr}\{\mathbf{\Omega}_\mathrm{z}^{+}\}$) but different elements, such that $\mathbf{Q}^{\prime}_\mathrm{z}$ in \eqref{e:QOptS1} is PSD if $\mathbf{\Omega}_\mathrm{z}^{+}$ in \eqref{e:QOptS1} is substituted by $\mathbf{\tilde{\Omega}}_\mathrm{z}^{+}$. Therefore, the power limit of the jammer as well as the gains of the eigen-channels determine whether or not $\mathbf{Q}_\mathrm{z}^{\prime}$ is PSD. The above fact, which reveals the effect of the jamming power limit and the jamming channel on the jammer's strategy, has not been observed before as the jamming channel has been neglected. While the simplified solution \eqref{e:QOptS1Special} and its scalar-form equivalence in \cite{JamMIMO3} can always be made PSD by forcing the negative elements to be zero and adjusting $\lambda$ to satisfy the power constraint, such method does not work for the model without neglecting the jamming channel as considered here. The problem of finding the solution when $\mathbf{Q}_\mathrm{z}^{\prime}$ in \eqref{e:QOptS1} is indefinite will be studied in Sections~\ref{s:numer}~and~\ref{s:subsolu}.

Now consider the general case that $\mathbf{H}_\mathrm{r}\mathbf{Q}_\mathrm{s}\mathbf{H}_\mathrm{r}^\mathrm{H}$ is PSD but not necessarily PD. Since $\mathbf{H}_\mathrm{r}\mathbf{Q}_\mathrm{s}\mathbf{H}_\mathrm{r}^\mathrm{H}$, or equivalently $\mathbf{B}$, is PSD but not necessarily PD in this case, $\mathbf{\tilde{B}}$ in \eqref{e:tildeB} and consequently $\mathbf{\tilde{A}}$ in \eqref{e:tildeA} can be rank deficient. In this situation, assume that the rank of $\mathbf{\tilde{A}}$ is $r_\mathbf{\tilde{A}}$ and denote the diagonal matrix made of the $r_\mathbf{\tilde{A}}$ positive eigenvalues of $\mathbf{\tilde{A}}$ as $\mathbf{\Lambda}_\mathbf{\tilde{A}}^{+}$. Let also the EVD of $\mathbf{\tilde{A}}$ be given as
\vspace{-1mm}
\begin{eqnarray}\label{e:AEVDblock}
\mathbf{\tilde{A}}\!\!\!\!&=&\!\!\!\! \mathbf{U}_{\mathbf{\tilde{A}}}\mathbf{\Lambda}_{\mathbf{\tilde{A}}}\mathbf{U}_{\mathbf{\tilde{A}}}^\mathrm{H} \nonumber\\
    &=&\!\!\!\!\!\!\!\!
    \kbordermatrix{\!\!
    ~ & \!\!r_\mathbf{\tilde{A}}\!\! & \!\!r_\mathrm{z}-r_\mathbf{\tilde{A}}\!\! \cr
      & {\mathbf{U}_{\mathbf{\tilde{A}}1}  } & \mathbf{U}_{\mathbf{\tilde{A}}2} }
          \!\!\!\!\kbordermatrix{\!\!
    ~   &  &  \cr
     & \mathbf{\Lambda}_\mathbf{\tilde{A}}^{+} \vspace{1mm} &
        \mathbf{0} \vspace{1mm} \cr
     &  \mathbf{0} &\!\!\mathbf{0} }
         \!\!\!\!\kbordermatrix{\!\!
    ~ & \cr
      & \mathbf{U}_{\mathbf{\tilde{A}}1}^\mathrm{H} \vspace{2mm}\cr
      & \mathbf{U}_{\mathbf{\tilde{A}}2}^\mathrm{H}}.
\end{eqnarray}

The following theorem regarding the solution in this general case is in order.

\textbf{Theorem~2}: When $\mathbf{H}_\mathrm{r}\mathbf{Q}_\mathrm{s}\mathbf{H}_\mathrm{r}^\mathrm{H}$ is PSD but not necessarily PD, the optimization problem \eqref{e:JOne} has the following closed-form optimal solution
\begin{eqnarray}\label{e:QOptS2}
\mathbf{Q}_\mathrm{z}^{\prime}\hspace{-3mm}&=&\hspace{-3mm}\mathbf{U}_{\mathbf{\tilde{A}}1}\!\sqrt{\!\frac{1}{\lambda}\mathbf{\Lambda}_\mathbf{\tilde{A}}^{+}\!\!+\!\!
\frac{1}{4}{\mathbf{\Lambda}_{\mathbf{\tilde{A}}}^{+}}^2\!}\mathbf{U}_{\mathbf{\tilde{A}}1}^\mathrm{H} \nonumber \\
&&\!- \frac{1}{2}\mathbf{U}_{\mathbf{\tilde{A}}1}\!\mathbf{\Lambda}_\mathbf{\tilde{A}}^{+}\!\mathbf{U}_{\mathbf{\tilde{A}}1}^\mathrm{H} \!-\sigma^2{\mathbf{\Omega}_\mathrm{z}^{+}}^{\!\!-1}\!{\mathbf{\Omega}_\mathrm{z}^{+}}^\mathrm{\!\!-H}
\end{eqnarray}
under the condition that the above matrix $\mathbf{Q}_\mathrm{z}^{\prime}$ is PSD, where $\lambda$ is chosen such that the jammer's power constraint \eqref{e:JOnecon} is satisfied with equality.

\textbf{Proof}: See Subsection~\ref{App:C5prfT2} in Appendix.

It can be seen that if $\mathbf{\tilde{A}}$ has full rank, then \eqref{e:QOptS2} is equivalent to \eqref{e:QOptS1}. Similarly, $\mathbf{Q}_\mathrm{z}^{\prime}$ given by \eqref{e:QOptS2} can be indefinite depending on the jammer's power limit $P_\mathrm{z}$ and the jamming channel $\mathbf{\Omega}_\mathrm{z}^{+}$. To tackle this problem, we next find solutions of the optimization problem \eqref{e:JOne} when $\mathbf{Q}_\mathrm{z}^{\prime}$ given in \eqref{e:QOptS1} or \eqref{e:QOptS2} is indefinite. We propose two different approaches. The first one is to find the optimal solution numerically. The second one is to find a suboptimal solution in closed-form. The two approaches provide a choice between accuracy and complexity. We start from describing an algorithm for finding the optimal solution of \eqref{e:JOne} numerically.

\section{Optimal numeric solution for single target jamming}\label{s:numer}

As mentioned earlier, the closed-form expressions for the matrix $\mathbf{Q}_\mathrm{z}^{\prime}$ given by \eqref{e:QOptS1} and \eqref{e:QOptS2}  when $\mathbf{H}_\mathrm{r}\mathbf{Q}_\mathrm{s}\mathbf{H}_\mathrm{r}^\mathrm{H}$ is PD and PSD, respectively, may not be valid when the power constraint $P_\mathrm{z}$ is small. Then, the optimal solution may not be found in closed-form.

Substituting \eqref{e:OmgzTt} and \eqref{e:Qtildet} into \eqref{e:JoneobjEqvt2} and using the definitions \eqref{e:tildeB} and \eqref{e:tildeA}, it can be shown\footnote{The details can be found in the proof of Theorem~1, from \eqref{e:JOneobjEqv2} to \eqref{e:RJEqv1}, Subsection~\ref{App:C5prfT1} in Appendix.} that the original problem of minimizing \eqref{e:JOneobjEqv1t} is equivalent to the minimization of
\begin{equation}\label{e:RJEqv3}
\bar{R}^\mathrm{J}=\log{\left|\mathbf{I}+\mathbf{\tilde{A}}(\mathbf{Q}^{\prime}_\mathrm{z}+\sigma^2{\mathbf{\Omega}_\mathrm{z}^{+}}^{-1}
{\mathbf{\Omega}_\mathrm{z}^{+}}^{-\mathrm{H}})^{-1}\right|}.
\end{equation}

Although the minimization of \eqref{e:RJEqv3} subject to the power constraint is a convex problem, it is not a disciplined convex problem \cite{CovDisciplin}. Therefore, the optimal solution cannot be obtained using classic convex optimization methods. In order to find the optimal solution, we first rewrite the problem into the following equivalent form
\begin{subequations}\label{e:LPOpteq}% low power case equivalent form
\begin{align}
&\hspace{-4mm} \mathop{\mathbf{min}}\limits_{\alpha, \mathbf{Q}^{\prime}_\mathrm{z}} \quad
\alpha - \log{|\mathbf{Q}^{\prime}_\mathrm{z}+ \mathbf{D}_0}| \label{e:LPOpteqobj}\\
&\hspace{-2mm} \mathbf{s.t.} \quad\, \alpha \geq \log{|\mathbf{Q}^{\prime}_\mathrm{z}+ \mathbf{D}_0+ \mathbf{\tilde{A}}|}   \label{e:LPOpteqcons1}\\
&\hspace{-2mm} \quad\quad\;\, \text{Tr}\{\mathbf{Q}^{\prime}_\mathrm{z}\}\leq P_{\rm z} \label{e:LPOpteqcons2}
\end{align}
\end{subequations}
in which $\mathbf{D}_0\triangleq\sigma^2{\mathbf{\Omega}_\mathrm{z}^{+}}^{-1}{\mathbf{\Omega}_\mathrm{z}^{+}}^\mathrm{\!-H}$. In the above problem, the objective function is convex while the first constraint is not. In order to solve the problem \eqref{e:LPOpteq}, we first consider the following problem in a similar form
\begin{subequations}\label{e:LPOptsub}
\begin{align}
&\hspace{-4mm} \mathop{\mathbf{min}}\limits_{\alpha, \mathbf{Q}^{\prime}_\mathrm{z}} \quad
\alpha - \log{|\mathbf{Q}^{\prime}_\mathrm{z}+ \mathbf{D}_0}| \label{e:LPOptsubobj}\\
&\hspace{-2mm} \mathbf{s.t.} \quad\, \alpha \geq \log{|{\mathbf{Q}^{\prime}}^{\dag}_\mathrm{z}\!+\!\mathbf{D}_0 \!+\! \mathbf{\tilde{A}}|} \!+\! \text{Tr}\{\big({\mathbf{Q}^{\prime}}^{\dag}_\mathrm{z} \!+\! \mathbf{D}_0  \nonumber\\
&\hspace{14mm} \!+\! \mathbf{\tilde{A}}\big)^{-1}\mathbf{Q}^{\prime}_\mathrm{z}\} - \text{Tr}\{\big({\mathbf{Q}^{\prime}}^{\dag}_\mathrm{z} \!+\! \mathbf{D}_0 \!+\! \mathbf{\tilde{A}}\big)^{-1}{\mathbf{Q}^{\prime}}^{\dag}_\mathrm{z}\} \label{e:LPOptsubcons1}\\
&\hspace{-2mm} \quad\quad\;\;\, \text{Tr}\{\mathbf{Q}^{\prime}_\mathrm{z}\}\leq P_{\rm z}. \label{e:LPOptsubcons2}
\end{align}
\end{subequations}
Here ${\mathbf{Q}^{\prime}}^{\dag}_\mathrm{z}$ stands for a given $\mathbf{Q}^{\prime}_\mathrm{z}$ subject to  \eqref{e:LPOpteqcons2}. The optimal solution of the problem \eqref{e:LPOpteq} can be found from solving the problem \eqref{e:LPOptsub} iteratively. Specifically, the corresponding algorithm is summarized in Table~\ref{t:OptbyAlg}. The following lemma regarding the algorithm in Table~\ref{t:OptbyAlg} is in order.
\begin{table}
\begin{center}
\caption {Steps for finding the optimal solution of the problem \eqref{e:LPOpteq}.}\label{t:OptbyAlg}
\begin{tabular*}{0.48\textwidth}[]{p{0.47\textwidth}}
\hline\hline
1. Select an initial matrix ${\mathbf{Q}^{\prime}}^{\dag}_\mathrm{z}$ subject to $\text{Tr}\{{\mathbf{Q}^{\prime}}^{\dag}_\mathrm{z}\}\leq P_{\rm z}$.\\
\hline 2. Solve the problem \eqref{e:LPOptsub} given ${\mathbf{Q}^{\prime}}^{\dag}_\mathrm{z}$. Denote the corresponding optimal solution of $\mathbf{Q}^{\prime}_\mathrm{z}$ as ${\mathbf{Q}^{\prime}}^{*}_\mathrm{z}$. \\
\hline 3. Set ${\mathbf{Q}^{\prime}}^{\dag}_\mathrm{z}={\mathbf{Q}^{\prime}}^{*}_\mathrm{z}$. \\
\hline 4. Repeat the Steps~2~and~3 until convergence.\\
\hline \hline
\end{tabular*}
\end{center}
\vspace{-0.6cm}
\end{table}

\emph{Lemma~3}: The matrix ${\mathbf{Q}^{\prime}}^{*}_\mathrm{z}$ in the procedure described in Table~\ref{t:OptbyAlg} converges to the optimal solution of the problem \eqref{e:LPOpteq}.

\textbf{Proof}: See Subsection~\ref{App:C5prfl3} in Appendix.

After obtaining the optimal ${\mathbf{Q}^{\prime}}^{*}_\mathrm{z}$ using the algorithm in Table~\ref{t:OptbyAlg}, the optimal matrix $\mathbf{Q}_\mathrm{z}$ can be obtained using \eqref{e:QOptform}.

\section{An alternative solution for single target jamming: suboptimal solution in closed-form}\label{s:subsolu}
The numerical method used for finding the optimal $\mathbf{Q}^{\prime}_\mathrm{z}$ in the previous section can be computationally complex as compared to obtaining a  closed-form solution. Therefore, we next give an approximation of the optimal solution in closed-form when the matrix $\mathbf{Q}_\mathrm{z}^{\prime}$ given by \eqref{e:QOptS1} (when $\mathbf{H}_\mathrm{r}\mathbf{Q}_\mathrm{s}\mathbf{H}_\mathrm{r}^\mathrm{H}$ is PD) or \eqref{e:QOptS2} (when $\mathbf{H}_\mathrm{r}\mathbf{Q}_\mathrm{s}\mathbf{H}_\mathrm{r}^\mathrm{H}$ is PSD) is indefinite.

When $\mathbf{H}_\mathrm{r}\mathbf{Q}_\mathrm{s}\mathbf{H}_\mathrm{r}^\mathrm{H}$ is PD, a suboptimal closed-form solution to the problem \eqref{e:JOne} when the matrix $\mathbf{Q}_\mathrm{z}^{\prime}$ in \eqref{e:QOptS1} is indefinite can be given as
\begin{equation}\label{e:QEst1}
\mathbf{Q}^{\prime}_\mathrm{z}=\mathbf{U}_{\mathbf{\tilde{A}}}\sqrt{\frac{1}{\tilde{\lambda}}\mathbf{\Lambda}_{\mathbf{\tilde{A}}}+
\frac{1}{4}\mathbf{\Lambda}_{\mathbf{\tilde{A}}}^2}\mathbf{U}_{\mathbf{\tilde{A}}}^{\rm H}-\frac{1}{2}\mathbf{\tilde{A}}+(\tilde{\epsilon}-1)\mathbf{D}_0
\end{equation}
where $\mathbf{D}_0$ is defined after \eqref{e:LPOpteq}, and $\tilde{\epsilon}$ and $\tilde{\lambda}$ are the optimal solution to the problem
\begin{subequations}\label{e:getParamets}
\begin{align}
&\hspace{-4mm} \mathop{\mathbf{min}}\limits_{\epsilon, \lambda} \quad \epsilon \\
&\hspace{-2mm} \mathbf{s.t.} \quad\; \mathbf{U}_{\mathbf{\tilde{A}}}\sqrt{\frac{1}{\lambda}\mathbf{\Lambda}_{\mathbf{\tilde{A}}}\!+\!
\frac{1}{4}\mathbf{\Lambda}_{\mathbf{\tilde{A}}}^2}\mathbf{U}_{\mathbf{\tilde{A}}}^{\rm H}\!-\!\frac{1}{2}\mathbf{\tilde{A}}\!+\!(\epsilon\!-\!1)\mathbf{D}_0\succeq 0 \label{e:getPcons1}\\
&\hspace{-2mm} \quad\quad\;\;\, \text{Tr}\left\{\sqrt{\frac{1}{\lambda}\mathbf{\Lambda}_{\mathbf{\tilde{A}}}\!+\!\frac{1}{4}\mathbf{\Lambda}_{\mathbf{\tilde{A}}}^2}
\!-\!\frac{1}{2}\mathbf{\tilde{A}}\!+\!(\epsilon\!-\!1)\mathbf{D}_0\right\}= P_{\rm z}\\
&\hspace{-2mm} \quad\quad\;\;\,  0 \leq \epsilon\leq 1 \\
&\hspace{-2mm} \quad\quad\;\;\,  \lambda>0. \label{e:getPcons4}
\end{align}
\end{subequations}

It is worth mentioning that the constraints \eqref{e:getPcons1}-\eqref{e:getPcons4} specify a non-empty feasible set. It can be seen that the suboptimal solution \eqref{e:QEst1} is equivalent to the expression in \eqref{e:QOptS1} plus the term $\tilde{\epsilon} \mathbf{D}_0$ (using the definitions \eqref{e:tildeA} and $\mathbf{D}_0\triangleq\sigma^2{\mathbf{\Omega}_\mathrm{z}^{+}}^{-1}{\mathbf{\Omega}_\mathrm{z}^{+}}^\mathrm{\!-H}$). The logic behind the suboptimal solution \eqref{e:QEst1} is that the remaining part of the expression \eqref{e:QOptS1} without $-\mathbf{D}_0$ is always PSD. Therefore, there exists a non-negative factor $\epsilon<1$ such that the summation is PSD if $-\mathbf{D}_0$ is scaled by $1-\epsilon$ and added back to the remaining part of \eqref{e:QOptS1}. In order to remain as close as possible to the form of \eqref{e:QOptS1} in the above modification, the minimum $\epsilon$ that results in a PSD $\mathbf{Q}_\mathrm{z}^{\prime}$ is used.

The above suboptimal solution given by \eqref{e:QEst1} is proposed based on the following reasons. First and most important, it can be shown that $\mathbf{Q}_\mathrm{z}^{\prime}$ given by the above suboptimal solution is the same as $\mathbf{Q}_\mathrm{z}^{\prime}$ given by \eqref{e:QOptS1} when the latter one is PSD (and consequently $\tilde{\epsilon}=0$). Therefore, the use of \eqref{e:QEst1} is sufficient for calculating the jamming strategy in all cases because \eqref{e:QEst1} gives the optimal solution when it exists in closed-form and gives the suboptimal solution otherwise. Second, when it is not optimal, the suboptimal solution given by \eqref{e:QEst1} is in fact very close to the optimal one found numerically (as will be shown in simulations). Third, compared to the numerical solution in Section~IV, the suboptimal solution given by \eqref{e:QEst1} can be obtained with negligible complexity since the parameters $\tilde{\epsilon}$ and $\tilde{\lambda}$ can be obtained by a simple bisectional search. Last, the above suboptimal solution is always PSD as can be seen from the constraint \eqref{e:getPcons1}.

The closed-form suboptimal solution for the general case when $\mathbf{H}_\mathrm{r}\mathbf{Q}_\mathrm{s}\mathbf{H}_\mathrm{r}^\mathrm{H}$ is PSD but not necessarily PD can be obtained similarly. In this case, the suboptimal solution is expressed in closed-form as 
\begin{eqnarray}\label{e:QprimeEstS2}
\mathbf{Q}^{\prime}_\mathrm{z}\hspace{-3mm}&=&\hspace{-3mm}\mathbf{U}_{\mathbf{\tilde{A}}1}\!\sqrt{\!\frac{1}{\tilde{\lambda}}\mathbf{\Lambda}_\mathbf{\tilde{A}}^{+}\!+\!
\frac{1}{4}{\mathbf{\Lambda}_{\mathbf{\tilde{A}}}^{+}}^2\!}\mathbf{U}_{\mathbf{\tilde{A}}1}^\mathrm{H} \nonumber \\
&&\!-  \frac{1}{2}\mathbf{U}_{\mathbf{\tilde{A}}1}\!\mathbf{\Lambda}_\mathbf{\tilde{A}}^{+}\!\mathbf{U}_{\mathbf{\tilde{A}}1}^\mathrm{H} \!+(\tilde{\epsilon}-1)\mathbf{D}_0
\end{eqnarray}
where $\tilde{\epsilon}$ and $\tilde{\lambda}$ are the optimal solution to the problem
\begin{subequations}\label{e:getParametsS2}
\begin{align}
&\hspace{-2mm} \mathop{\mathbf{min}}\limits_{\epsilon, \lambda} \; \epsilon \\
&\hspace{-1mm} \mathbf{s.t.} \;\; \mathbf{U}_{\mathbf{\tilde{A}}1}\!\sqrt{\!\frac{1}{\lambda}\mathbf{\Lambda}_\mathbf{\tilde{A}}^{\!+}\!\!+\!
\frac{1}{4}{\mathbf{\Lambda}_{\mathbf{\tilde{A}}}^{\!+}}^2\!}\mathbf{U}_{\mathbf{\tilde{A}}1}^\mathrm{H} \!\nonumber \\
&\hspace{6mm} \quad\;\;\;\,- \frac{1}{2}\mathbf{U}_{\mathbf{\tilde{A}}1}\!\mathbf{\Lambda}_\mathbf{\tilde{A}}^{\!+}\!\mathbf{U}_{\mathbf{\tilde{A}}1}^\mathrm{H} \!\!+\!(\epsilon\!\!-\!1)\mathbf{D}_0\!\succeq\! 0 \label{e:getPcons1r}\\
&\hspace{-2mm} \quad\;\;\;\, \text{Tr}\left\{\sqrt{\!\frac{1}{\lambda}\mathbf{\Lambda}_\mathbf{\tilde{A}}^{+}\!+\!
\frac{1}{4}{\mathbf{\Lambda}_{\mathbf{\tilde{A}}}^{+}}^2\!}- \frac{1}{2}\mathbf{\Lambda}_\mathbf{\tilde{A}}^{+}
+(\epsilon-1)\mathbf{D}_0\right\}= P_{\rm z}\\
&\hspace{-2mm} \quad\;\;\;\,  0 \leq \epsilon\leq 1 \\
&\hspace{-2mm} \quad\;\;\;\,  \lambda>0. \label{e:getPcons4r}
\vspace{-1mm}
\end{align}
\end{subequations}

\begin{table}
\begin{center}
\caption {Summary of the procedure for finding the solution to the optimization problem \eqref{e:JOne}.}\label{t:JamOneCPLAlg}
\begin{tabular*}{0.49\textwidth}[]{p{0.47\textwidth}}
\hline\hline 1. Check whether or not $\mathbf{H}_\mathrm{r}\mathbf{Q}_\mathrm{s}\mathbf{H}_\mathrm{r}^\mathrm{H}$ is PD. If yes, obtain $\mathbf{Q}_\mathrm{z}^{\prime}$ using \eqref{e:QOptS1}. Otherwise, obtain $\mathbf{Q}_\mathrm{z}^{\prime}$ using \eqref{e:QOptS2}.\\
\hline 2. Check whether or not the above obtained $\mathbf{Q}_\mathrm{z}^{\prime}$ is PSD. If yes, substitute the obtained $\mathbf{Q}_\mathrm{z}^{\prime}$ into \eqref{e:QOptform} to find the optimal $\mathbf{Q}_\mathrm{z}$. Otherwise, select from two options: a) finding optimal numerical solution; b) finding suboptimal closed-form solution. For a), proceed to step~3. For b), proceed to  step~4. \\
\hline 3. Use the algorithm in Table~\ref{t:OptbyAlg} to obtain the optimal numerical solution. Exit.\\
\hline 4. Obtain $\tilde{\epsilon}$ and $\tilde{\lambda}$ by solving the problem \eqref{e:getParamets} (if $\mathbf{H}_\mathrm{r}\mathbf{Q}_\mathrm{s}\mathbf{H}_\mathrm{r}^\mathrm{H}$ is PD) or problem \eqref{e:getParametsS2} (if $\mathbf{H}_\mathrm{r}\mathbf{Q}_\mathrm{s}\mathbf{H}_\mathrm{r}^\mathrm{H}$ is PSD but not PD). Then obtain the suboptimal closed-form solution by using \eqref{e:QOptform} with \eqref{e:QEst1} (if $\mathbf{H}_\mathrm{r}\mathbf{Q}_\mathrm{s}\mathbf{H}_\mathrm{r}^\mathrm{H}$ is PD) or \eqref{e:QprimeEstS2} (if $\mathbf{H}_\mathrm{r}\mathbf{Q}_\mathrm{s}\mathbf{H}_\mathrm{r}^\mathrm{H}$ is PSD but not PD). Exit.\\
\hline \hline
\end{tabular*}
\end{center}
\vspace{-0.6cm}
\end{table}

With the proposed closed-form optimal and suboptimal solutions and the algorithm for finding the optimal numerical solution, the complete procedure of calculating the jamming strategy $\mathbf{Q}_\mathrm{z}$ is summarized in Table~\ref{t:JamOneCPLAlg}.

\section{Discussion: extension to multiple legitimate signals}\label{s:mutlijam}
The investigation on the jamming strategy in the preceding sections focuses on the case of jamming one legitimate signal between a single transceiver pair. However, it is possible to extend the previously obtained results to the case of jamming multiple legitimate signals with the objective of minimizing the sum-rate of the legitimate communications. In this section, we consider several different scenarios with multiple legitimate signals and briefly investigate the jamming strategies in these scenarios based on the previous results.

\subsection{Multiple legitimate signals on a multiple access channel}

The extension of the jamming strategy to jamming multiple signals in the scenario of multiple access channel is immediate. Assume that there are $m$ legitimate transmitters sending signals to a common receiver. Denote the covariance of the $i$th legitimate signal as $\mathbf{Q}_i$ and the channel from the $i$th legitimate transmitter to the receiver as $\mathbf{H}_i$. With $\mathbf{Q}_\mathrm{z}$ denoting the covariance matrix of the jamming signal and $\mathbf{H}_\mathrm{z}$ denoting the jamming channel, the sum-rate of the multiple access channel under jamming can be written as
\begin{equation}
R_\mathrm{ma}^{\rm J}=\log\left|\mathbf{I}+\sum\limits_i\mathbf{H}_i\mathbf{Q}_i\mathbf{H}_i^\mathrm{H}(\mathbf{H}_\mathrm{z}
\mathbf{Q}_\mathrm{z}\mathbf{H}_\mathrm{z}^\mathrm{H} +\sigma^2\mathbf{I})^{-1}\right|.
\end{equation}
It can be seen that the results on the closed-form solution in Theorems~1~and~2, the numerical method described in Table~\ref{t:OptbyAlg}, and the results on the closed-form suboptimal solutions given by \eqref{e:QEst1} and \eqref{e:QprimeEstS2} are also valid if the term $\mathbf{H}_\mathrm{r}\mathbf{Q}_\mathrm{s}\mathbf{H}_\mathrm{r}^\mathrm{H}$ is substituted by $\sum\limits_i\mathbf{H}_i\mathbf{Q}_i\mathbf{H}_\mathrm{i}^\mathrm{H}$.

\subsection{Multiple legitimate signals on a broadcast channel}

Assume that a legitimate transmitter (base station) is broadcasting to $m$ receivers. Denote the covariance of the legitimate signal as $\mathbf{Q}_\mathrm{s}$ and the channel from the legitimate transmitter to the $i$th receiver as $\mathbf{H}_i$. The noise covariance at receiver $i$ is denoted as $\sigma_i^2\mathbf{I}$. With $\mathbf{Q}_\mathrm{z}$ denoting the covariance matrix of the jamming signal and $\mathbf{H}_{\mathrm{z}i}$ denoting the jamming channel from the jammer to the $i$th legitimate receiver, the sum-rate of the broadcast channel in the presence of jamming can be written as\cite{MIMOBC}
\begin{equation}
R_\mathrm{bc}^{\rm J} = \log{|\mathbf{H}\mathbf{Q}_\mathrm{s}\mathbf{H}^\mathrm{H}+ \mathbf{D} +\mathbf{\Theta}_\mathrm{z}|} -\log{|\mathbf{D}+\mathbf{\Theta}_\mathrm{z}|}
\end{equation}
where $\mathbf{H}=[\mathbf{H}_1^{\mathrm{H}}, \dots, \mathbf{H}_m^{\mathrm{H}}]^{\mathrm{H}}$, $\mathbf{D}$ is a diagonal matrix with its $i$th ($i=1, \dots, m$) diagonal block being $\sigma_i^2\mathbf{I}$, and $\mathbf{\Theta}_\mathrm{z}$ is
a PSD matrix with the $i$th ($i=1, \dots, m$) diagonal block given as $\mathbf{H}_{\mathrm{z}i}\mathbf{Q}_\mathrm{z}\mathbf{H}_{\mathrm{z}i}^{\rm H}$. The size of the $i$th ($i=1, \dots, m$) diagonal block of both $\mathbf{D}$ and $\mathbf{\Theta}_\mathrm{z}$ is equal to the number of antennas at the $i$th ($i=1, \dots, m$) receiver.

In this scenario, the closed-form expressions derived for the optimal and sub-optimal jamming strategies in Sections~III~and~V are not applicable anymore. However, the numeric method used in Table~\ref{t:OptbyAlg} can still be applied to obtain the optimal jamming solution after slight changes. Specifically, the problem of minimizing $R_\mathrm{bc}^{\rm J}$ in this scenario can be rewritten into the following form
\begin{subequations}\label{e:MuBC}% low power case equivalent form
\begin{align}
&\hspace{-6mm} \mathop{\mathbf{min}}\limits_{\alpha, \mathbf{\Theta}_\mathrm{z}, \mathbf{Q}_\mathrm{z}} \quad
\alpha -\log{|\mathbf{D}+\mathbf{\Theta}_\mathrm{z}|} \label{e:MuBCobj}\\
&\hspace{-1mm} \mathbf{s.t.} \quad\;\, \alpha \geq \log{|\mathbf{H}\mathbf{Q}_\mathrm{s}\mathbf{H}^\mathrm{H}+ \mathbf{D} +\mathbf{\Theta}_\mathrm{z}|}  \label{e:MuBCcons1}\\
&\hspace{-1mm} \quad\quad\;\;\, \mathbf{\Theta}_\mathrm{z}^{(i)} = \mathbf{H}_{\mathrm{z}i}\mathbf{Q}_\mathrm{z}\mathbf{H}_{\mathrm{z}i}^{\rm H}, \forall i \label{e:MuBCcons2}\\
&\hspace{-1mm} \quad\quad\;\;\, \text{Tr}\{\mathbf{Q}_\mathrm{z}\}\leq P_{\rm z} \label{e:MuBCcons3}
\end{align}
\end{subequations}
where $\mathbf{\Theta}_\mathrm{z}^{(i)}$ denotes the $i$th ($i=1, \dots, m$) block on the diagonal of $\mathbf{\Theta}_\mathrm{z}$. Similar to the case of single-target jamming, the solution of the problem~\eqref{e:MuBC} can be found from solving the following problem
\begin{subequations}\label{e:MuBCsub}
\begin{align}
&\hspace{-3mm} \mathop{\mathbf{min}}\limits_{\alpha, \mathbf{\Theta}_\mathrm{z}, \mathbf{Q}_\mathrm{z}} \;\;
\alpha -\log{|\mathbf{D}+\mathbf{\Theta}_\mathrm{z}|} \label{e:MuBCsubobj}\\
&\hspace{1mm} \mathbf{s.t.} \quad\;\, \alpha \geq \log{|\mathbf{H}\mathbf{Q}_\mathrm{s}\mathbf{H}^\mathrm{H} \!\!+\! \mathbf{D} \!+\! \mathbf{\Theta}_\mathrm{z}^\dag|} \!+\! \text{Tr}\{\big(\mathbf{H}\mathbf{Q}_\mathrm{s}\mathbf{H}^\mathrm{H} \! \nonumber\\
&\hspace{16mm} \!+\! \mathbf{D} \!+\! \mathbf{\Theta}_\mathrm{z}^\dag\big)^{\!-1}(\mathbf{\Theta}_\mathrm{z}-\mathbf{\Theta}_\mathrm{z}^\dag)\}  \label{e:MuBCsubcons1}\\
&\hspace{2mm} \quad\quad\;\;\, \mathbf{\Theta}_\mathrm{z}^{(i)} = \mathbf{H}_{\mathrm{z}i}\mathbf{Q}_\mathrm{z}\mathbf{H}_{\mathrm{z}i}^{\rm H}, \forall i \label{e:MuBCsubcons2}\\
&\hspace{2mm} \quad\quad\;\;\, \text{Tr}\{\mathbf{Q}_\mathrm{z}\}\leq P_{\rm z} \label{e:MuBCsubcons3}
\end{align}
\end{subequations}
where $\mathbf{\Theta}_\mathrm{z}^{\dag}$ stands for a given $\mathbf{\Theta}_\mathrm{z}$ subject to  \eqref{e:MuBCcons2} and \eqref{e:MuBCcons3}. The optimal solution of the problem \eqref{e:MuBC} can be found by solving the problem \eqref{e:MuBCsub}, updating $\mathbf{\Theta}_\mathrm{z}^{\dag}$ using the optimal solution, and then solving the problem \eqref{e:MuBCsub} with the updated $\mathbf{\Theta}_\mathrm{z}^{\dag}$ until convergence.

\subsection{Multiple transceiver pairs with orthogonal transmissions}\label{ss:MultOrtho}
Now consider a system with $m$ legitimate transceiver pairs in which the transmissions of the $m$ legitimate signals are orthogonal, e.g., based on time division multiplexing (TDM) or frequency division multiplexing (FDM). Here we use TDM as an example. Denote the total transmission time duration of all legitimate signals as $t$ and the transmission time duration of the $i$th legitimate signal as $t_i$. Denote the covariance of the $i$th legitimate signal as $\mathbf{Q}_i$ and the channel between the $i$th legitimate transceiver pair as $\mathbf{H}_i$. The noise covariance at receiver $i$ is denoted as $\sigma_i^2\mathbf{I}$. The covariance matrix of the jamming signal in the $i$th interval (i.e., the transmission time duration of the $i$th legitimate signal) is $\mathbf{Q}_{\mathrm{z}i}$ and the channel from the jammer to the $i$th receiver is  $\mathbf{H}_{\mathrm{z}i}$. Given that the signal transmissions are orthogonal, the optimal $\mathbf{Q}_{\mathrm{z}i}$ only depends on $\mathbf{Q}_i$, $\mathbf{H}_i$, and $\mathbf{H}_{\mathrm{z}i}$. Therefore, previous results on the closed-form expression, numeric method, and suboptimal solution for single-target jamming could be applied here for each target signal. The difference is that the optimal power allocation for jamming the $m$ target signals needs to be determined for the scenario of multiple transceiver pairs with orthogonal transmissions. The sum-rate under jamming is expressed as
\begin{equation}
R_\mathrm{ot}^\mathrm{J}=\sum\limits_i \beta_i \! \log\!|\mathbf{I}\!+\!\mathbf{H}_i\mathbf{Q}_i\mathbf{H}_i^\mathrm{H}(\mathbf{H}_{\mathrm{z}i}
\mathbf{Q}_{\mathrm{z}i}\mathbf{H}_{\mathrm{z}i}^\mathrm{H}\!+\!\sigma_i^2\mathbf{I})^{\!-1}\!|
\end{equation}
where $\beta_i=t_i/t, i=1,\dots, m$. The problem in this scenario can be formulated as
\begin{subequations}\label{e:JMultPair} 
\begin{align}
&\hspace{-2mm} \mathop{\mathbf{min}}\limits_{\mathbf{Q}_{\mathrm{z}i}, \forall i}  \;R_\mathrm{ot}^\mathrm{J} \!  \label{e:obj} \\
&\hspace{-1mm} \mathbf{\;s.t.}  \;\;\,  \sum\limits_i \beta_i\text{Tr}\{\mathbf{Q}_{\mathrm{z}i}\}\leq P_\mathrm{z}. \label{e:con}
\end{align}
\end{subequations}

Assume that the proportion of the total jamming power $P_\mathrm{z}$ allocated for jamming the $i$th target signal is $\rho_i$.
When the number of legitimate transceiver pairs is small and the total transmission time is uniformly divided among all legitimate communications (i.e., $t_1=t_2\dots=t_m$), the problem can be solved by performing a search over the combinations $\{\rho_1, \rho_2, \dots, \rho_m\}$'s. For each combination $\{\rho_1, \rho_2, \dots, \rho_m\}$, the jamming strategy of each transceiver pair can be found using the previous results on the closed-form optimal/suboptimal solutions in Sections~III~and~V or the algorithm in Table~\ref{t:OptbyAlg} in Section~IV, with $\mathbf{Q}_\mathrm{s}$, $\mathbf{H}_\mathrm{r}$, $\mathbf{H}_{\mathrm{z}}$, $\mathbf{Q}_\mathrm{z}^{\prime}$, and $\mathbf{Q}_\mathrm{z}$  replaced by $\mathbf{Q}_i$, $\mathbf{H}_i$, $\mathbf{H}_{\mathrm{z}i}$, $\mathbf{Q}_{\mathrm{z}i}^{\prime}$, and $\mathbf{Q}_{\mathrm{z}i}$, respectively.

When the number of legitimate transceiver pairs is large or total transmission time is not uniformly divided, then previous results on closed-form solutions may not be applied. However, the method used for deriving the numerical solution can be used after a slight modification. The problem \eqref{e:JMultPair} is equivalent to the following problem
\begin{subequations}\label{e:JMultPair_Eq2} % Jam multiple transceiver pairs
\begin{align}
&\hspace{-2mm} \mathop{\mathbf{min}}\limits_{\mathbf{Q}_{\mathrm{z}i}, \alpha_i, \forall i}  \; \sum\limits_i \beta_i (\alpha_i\!- \! \log|\mathbf{H}_{\mathrm{z}i}\mathbf{Q}_{\mathrm{z}i}\mathbf{H}_{\mathrm{z}i}^\mathrm{H}\!+\!\sigma_i^2\mathbf{I}|)\!  \label{e:obj} \\
&\hspace{-1mm} \mathbf{\;s.t.}  \;\;\, \alpha_i\geq \log| \mathbf{H}_i\mathbf{Q}_i\mathbf{H}_i^\mathrm{H}+ \mathbf{H}_{\mathrm{z}i}\mathbf{Q}_{\mathrm{z}i}\mathbf{H}_{\mathrm{z}i}^\mathrm{H}\!+\!\sigma_i^2\mathbf{I} |, \forall i \\
&\hspace{6mm}  \;\;\,\sum\limits_i \beta_i\text{Tr}\{\mathbf{Q}_{\mathrm{z}i}\}\leq P_\mathrm{z}.  \label{e:con}
\end{align}
\end{subequations}
Following the idea of defining the equivalent jamming channel and equivalent jamming covariance given by the equations \eqref{e:OmgzTt}-\eqref{e:QOptform}, the solution to the above problem can be found similarly as finding the solution to \eqref{e:LPOpteq} through solving a subproblem similar to \eqref{e:LPOptsub}.

\subsection{Multiple transceiver pairs with interference}
It is also possible that there are $m$ legitimate transceiver pairs with transmissions spread over the same time interval and frequency band. Thereby the legitimate transmissions interfere with each other. Unlike the scenario with orthogonal transmissions, the jammer has only one jamming covariance to optimize instead of $m$ in Subsection~\ref{ss:MultOrtho}. Following the definitions of all channels and the legitimate signal covariances in Subsection~\ref{ss:MultOrtho}, the sum-rate under jamming in this scenario is given as
\begin{eqnarray}
&&\hspace{-1.5cm} R_\mathrm{ic}^{\rm J}=\sum\limits_i \log\big|\mathbf{I}+\mathbf{H}_i\mathbf{Q}_i\mathbf{H}_i^\mathrm{H}\big(\mathbf{H}_{\mathrm{z}i}
\mathbf{Q}_\mathrm{z}\mathbf{H}_{\mathrm{z}i}^\mathrm{H} \nonumber \\
&&\hspace{1.5cm} +\sum\limits_{j\neq i} \mathbf{H}_{ji}\mathbf{Q}_j\mathbf{H}_{ji}^\mathrm{H} +\sigma_i^2\mathbf{I}\big)^{-1}\big|
\end{eqnarray}
where $\mathbf{H}_{ji}$ represents the interference channel from the transmitter of the $j$th transceiver pair to the receiver of the $i$th transceiver pair.

The minimization of the above sum-rate is formulated as
\begin{subequations}\label{e:JMultPair_Eq3} % Jam multiple transceiver pairs
\begin{align}
&\hspace{-2mm} \mathop{\mathbf{min}}\limits_{\mathbf{Q}_{\mathrm{z}i}, \alpha_i, \forall i}  \; \sum\limits_i \big(\alpha_i\!- \! \log\big|\mathbf{H}_{\mathrm{z}i}\mathbf{Q}_\mathrm{z}\mathbf{H}_{\mathrm{z}i}^\mathrm{H}\!+\!\sum\limits_{j\neq i} \mathbf{H}_{ji}\mathbf{Q}_j\mathbf{H}_{ji}^\mathrm{H}\!+\!\sigma_i^2\mathbf{I}\big|\big)\!  \label{e:obj} \\
&\hspace{-1mm} \mathbf{\;s.t.}  \;\;\, \alpha_i\geq \log\big| \mathbf{H}_i\mathbf{Q}_i\mathbf{H}_i^\mathrm{H}\!+\! \mathbf{H}_{\mathrm{z}i}\mathbf{Q}_{\mathrm{z}i}\mathbf{H}_{\mathrm{z}i}^\mathrm{H}\!\nonumber \\
&\hspace{3.5cm}+\!\sum\limits_{j\neq i} \mathbf{H}_{ji}\mathbf{Q}_j\mathbf{H}_{ji}^\mathrm{H}\!+\!\sigma_i^2\mathbf{I} \big|, \forall i \\
&\hspace{6mm}  \;\;\, \text{Tr}\{\mathbf{Q}_\mathrm{z}\}\leq P_\mathrm{z}.  \label{e:con}
\end{align}
\end{subequations}
In this scenario, the previous results on the closed-form expressions in Sections~\ref{C5ss:OneCloseF}~and~\ref{s:subsolu}  are not applicable. The numerical solution to the above problem can be found similarly to finding the solutions to the problems \eqref{e:LPOpteq} and \eqref{e:JMultPair_Eq2}. The details are omitted due to the similarity.

\section{Simulations}\label{s:simula}
In this section, we demonstrate the obtained results on the jamming strategies for the cases of single target signal and multiple target signals. For multi-target jamming, we select the scenarios of broadcast channel and multiple transceivers with orthogonal transmissions as examples for the following reasons. First, the problem of multi-target jamming in multiple access channel is a straightforward extension of the single target jamming problem. Second, the scenario of multiple transceiver pairs with interference is solved in Section VI-D using the numerical method similar to the one used in the scenario of multiple transceiver pairs with orthogonal transmissions.

\begin{figure}[!t]
\begin{center}
%\hbox{\hspace{-1.2em}\includegraphics[angle=0,width=0.55\textwidth]{mainJ_RjverusPz.eps}}% for two-column
\includegraphics[angle=0,width=0.85\textwidth]{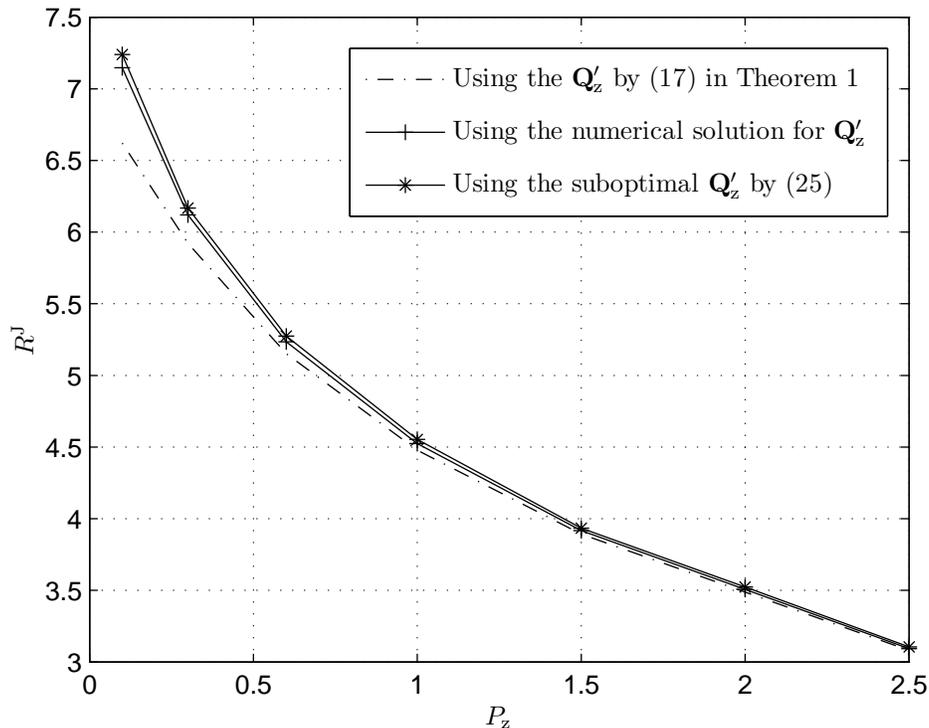}% for one-column
\end{center}
\vspace{-10mm}
\caption {Comparison of $R^\mathrm{J}$ versus $P_\mathrm{z}$ with $\mathbf{Q}_\mathrm{z}^{\prime}$ given by \eqref{e:QOptS1}, the optimal numerical solution, and \eqref{e:QEst1}, respectively.} \label{f:RjverusPz}
\vspace{-2mm}
\end{figure}

\begin{figure}[!t]
\begin{center}
%\hbox{\hspace{-1.2em}\includegraphics[angle=0,width=0.54\textwidth]{mainJ_PSDpercent.eps}}%
\includegraphics[angle=0,width=0.84\textwidth]{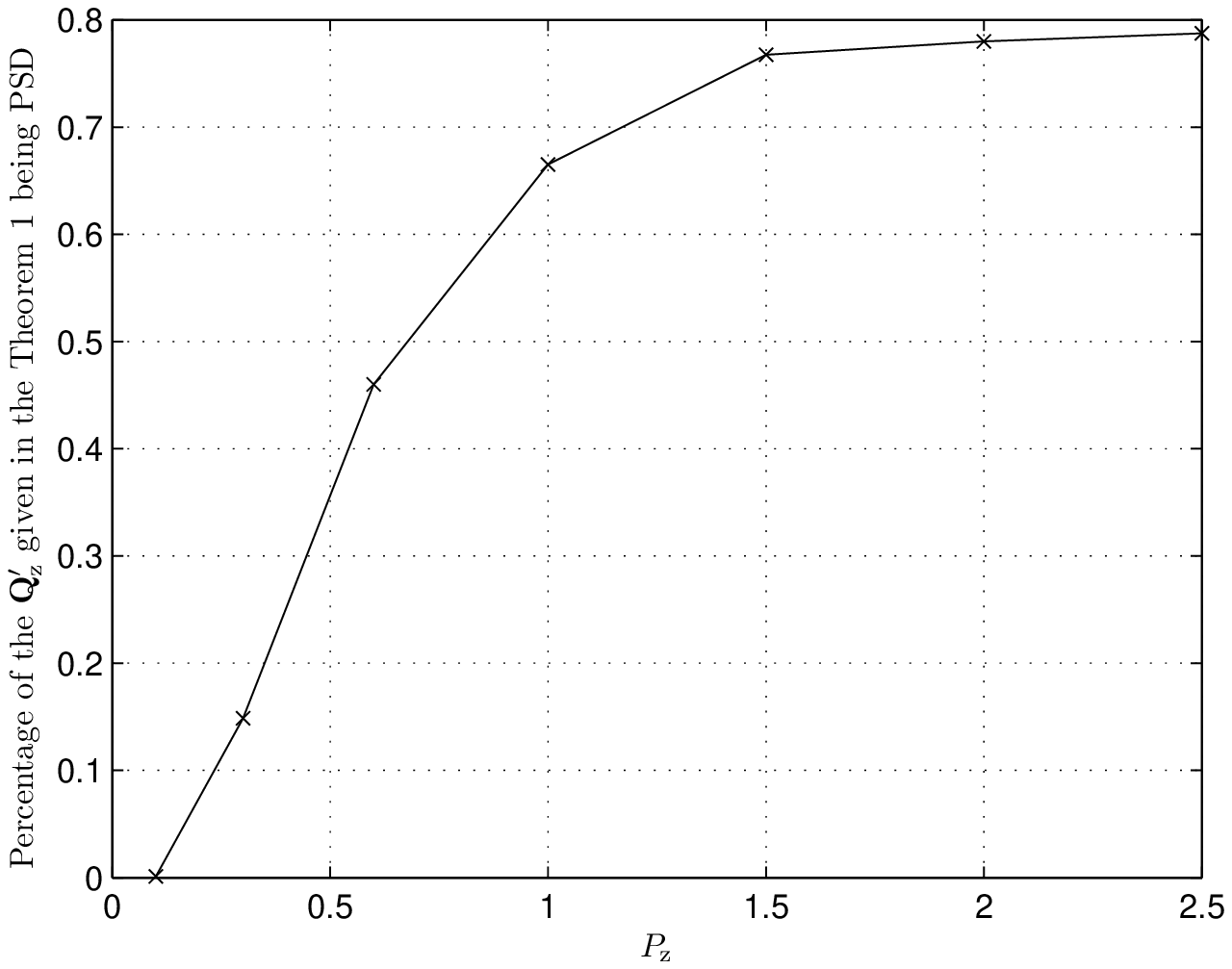}%
\end{center}
\vspace{-10mm}
\caption {Percentage that $\mathbf{Q}_\mathrm{z}^{\prime}$ given by \eqref{e:QOptS1} is PSD versus $P_\mathrm{z}$.} \label{f:PSDpercent}
\vspace{-2mm}
\end{figure}

\emph{Example~1: The case of a single target signal.} In this example, we compare the rates of the legitimate communication under jamming when the jammer's strategy $\mathbf{Q}_\mathrm{z}^{\prime}$ is given by (i) the expression in \eqref{e:QOptS1}, (ii) the optimal solution obtained numerically, and (iii) the approximation in \eqref{e:QEst1}, respectively.

The specific setup of this example is as follows. The number of antennas at the legitimate transmitter and receiver are set to be 4 and 3, respectively, while the number of antennas at the jammer is $5$. The transmit power for the legitimate transmitter is $3$ and the power allocation at the legitimate transmitter is based on waterfilling. The noise variance $\sigma^2$ is set to be 1. The elements of the channels $\mathbf{H}_{\mathrm{r}}$ and $\mathbf{H}_{\mathrm{z}}$ are generated from complex Gaussian distribution with zero mean and unit variance. As a result $\mathbf{H}_\mathrm{r}\mathbf{Q}_\mathrm{s}\mathbf{H}_\mathrm{r}^\mathrm{H}$ is always PD. We use 800 channel realizations and calculate the average $R^\mathrm{J}$ versus the power limit of the jammer $P_\mathrm{z}$.

Fig.~\ref{f:RjverusPz} shows the average $R^\mathrm{J}$ with  $\mathbf{Q}_\mathrm{z}^{\prime}$ obtained using the three aforementioned methods. Three observations can be made from this figure. First, when $P_\mathrm{z}$ is small, there is a gap between the average $R^\mathrm{J}$ with $\mathbf{Q}_\mathrm{z}^{\prime}$ given by \eqref{e:QOptS1} and the average $R^\mathrm{J}$ with the optimal $\mathbf{Q}_\mathrm{z}^{\prime}$ found numerically. The gap exists because $\mathbf{Q}_\mathrm{z}^{\prime}$ given by \eqref{e:QOptS1} is not always PSD and when it is not PSD, it no longer gives the optimal solution of the problem. Second, the gap between the average $R^\mathrm{J}$ with $\mathbf{Q}_\mathrm{z}^{\prime}$ obtained numerically and the average $R^\mathrm{J}$ given by the suboptimal $\mathbf{Q}_\mathrm{z}^{\prime}$ in \eqref{e:QEst1} is very small. It verifies that the proposed suboptimal solution is in fact very close to the optimal solution of the considered problem. Third, the three curves of average $R^\mathrm{J}$ converge when $P_\mathrm{z}$ increases.

Fig.~\ref{f:PSDpercent} shows the percentage of times that the matrix $\mathbf{Q}_\mathrm{z}^{\prime}$ given by \eqref{e:QOptS1} is PSD in all 800 channel realizations. It verifies the aforementioned fact that $\mathbf{Q}_\mathrm{z}^{\prime}$ given by \eqref{e:QOptS1} can be indefinite when the jammer's power limit $P_\mathrm{z}$ is small. Even when $P_\mathrm{z}$ is larger (above 2), there remains a $20\%$ chance that $\mathbf{Q}_\mathrm{z}^{\prime}$ given by \eqref{e:QOptS1} is indefinite. This verifies the other fact that whether or not $\mathbf{Q}_\mathrm{z}^{\prime}$ given by \eqref{e:QOptS1} is PSD also depends on the jamming channel.

Using the observations from the above two figures, it can be concluded that the suboptimal solution given by \eqref{e:QEst1} is a very good approximation of the optimal jamming strategy since it is very close to the optimal one when $\mathbf{Q}_\mathrm{z}^{\prime}$ given by \eqref{e:QOptS1} is indefinite while it becomes optimal when $\mathbf{Q}_\mathrm{z}^{\prime}$ given by \eqref{e:QOptS1} is PSD.

\begin{figure}[!t]
\begin{center}
%\hbox{\hspace{-1.5em}\includegraphics[angle=0,width=0.54\textwidth]{NCjamBC.eps}}% for two columns
\includegraphics[angle=0,width=0.84\textwidth]{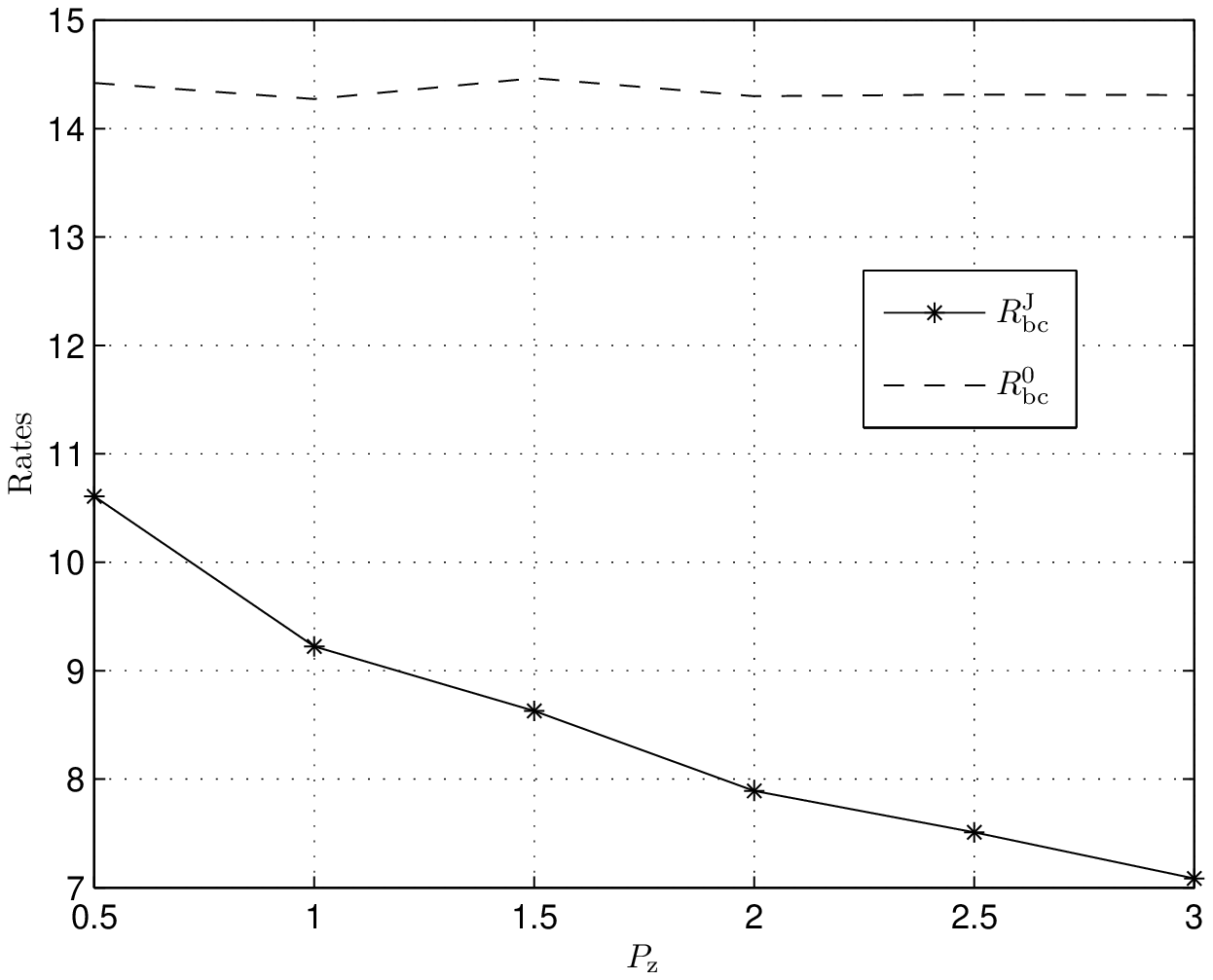}% for one column
\end{center}
\vspace{-10mm}
\caption {The sum-rate without jamming ($R_\mathrm{bc}^0$) and the sum-rate under optimal jamming ($R_\mathrm{bc}^\mathrm{J}$) versus $P_\mathrm{z}$ on a broadcast channel with one legitimate transmitter and three receivers.} \label{f:NCjamBC}
\vspace{-2mm}
\end{figure}

\emph{Example~2: Jamming multiple legitimate signals on a broadcast channel.} A broadcast channel with one legitimate transmitter and three legitimate receivers is considered. The specific setup of this simulation is as follows. The number of antennas at the legitimate transmitter is 4 while the numbers of antennas at the 1st, 2nd, and 3rd receivers are 3, 4, and 4, respectively. The number of antennas at the jammer is $4$. The transmit power for the legitimate transmitter is $3$. The noise variance $\sigma_i^2$ at the $i$th receiver is 0.5 for $i=1,2$ and 1 for $i=3$. The signal covariance $\mathbf{Q}_\mathrm{s}$ is assumed to be $\mathbf{I}$. The elements of the channels $\mathbf{H}_{i}$ and $\mathbf{H}_{\mathrm{z}i}, \forall i$ are generated from complex Gaussian distribution with zero mean and unit variance. We use 400 channel realizations and calculate the average $R_\mathrm{bc}^\mathrm{J}$ (obtained by iteratively solving \eqref{e:MuBCsub}) versus the power limit of the jammer $P_\mathrm{z}$. The sum-rate without jamming, denoted as $R_\mathrm{bc}^0$, is also calculated and averaged over the 400 channel realizations.

The above two sum-rates are shown in Fig.~\ref{f:NCjamBC}. From this figure, it can be seen that while $R_\mathrm{bc}^0$ is approximately a constant, the gap between $R_\mathrm{bc}^0$ and $R_\mathrm{bc}^\mathrm{J}$ evidently increases as $P_\mathrm{z}$ becomes larger. Thus, it shows that the jamming strategy used for the broadcast channel is efficient.

\begin{figure}[!t]
\begin{center}
%\hbox{\hspace{-1.2em}\includegraphics[angle=0,width=0.54\textwidth]{NCjamMUltiORT4001.eps}} %for twocolumns
\includegraphics[angle=0,width=0.84\textwidth]{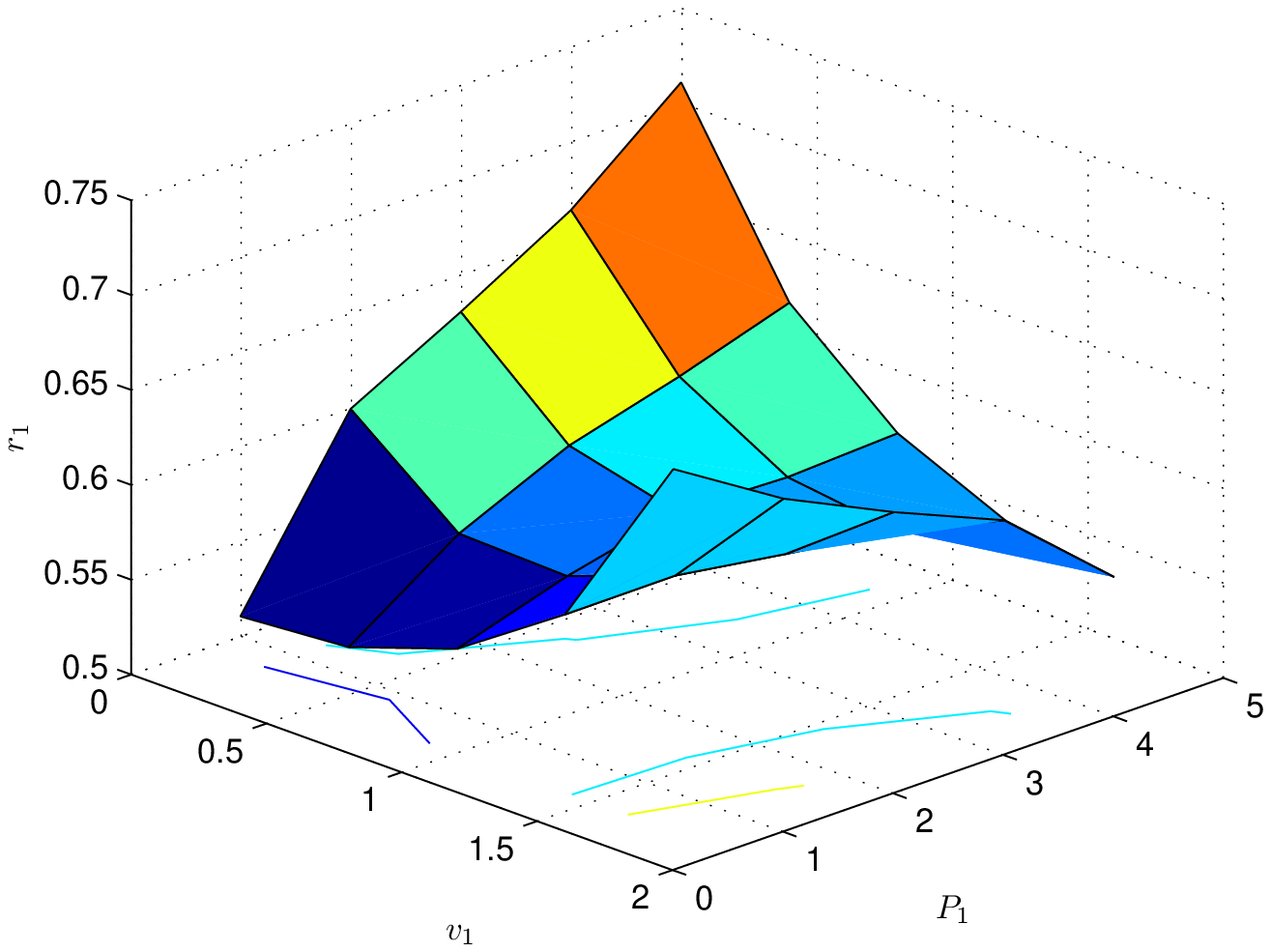} %for onecolumn
\end{center}
\vspace{-10mm}
\caption {The ratio of sum-rate reduction under the optimal jamming versus $P_1$ and $v_1$.} \label{f:NCjamMUltiORT4001}
\vspace{-2mm}
\end{figure}

\begin{figure}[!t]
\begin{center}
%\hbox{\hspace{-1.2em}\includegraphics[angle=0,width=0.54\textwidth]{NCjamMUltiORT4002.eps}}
\includegraphics[angle=0,width=0.84\textwidth]{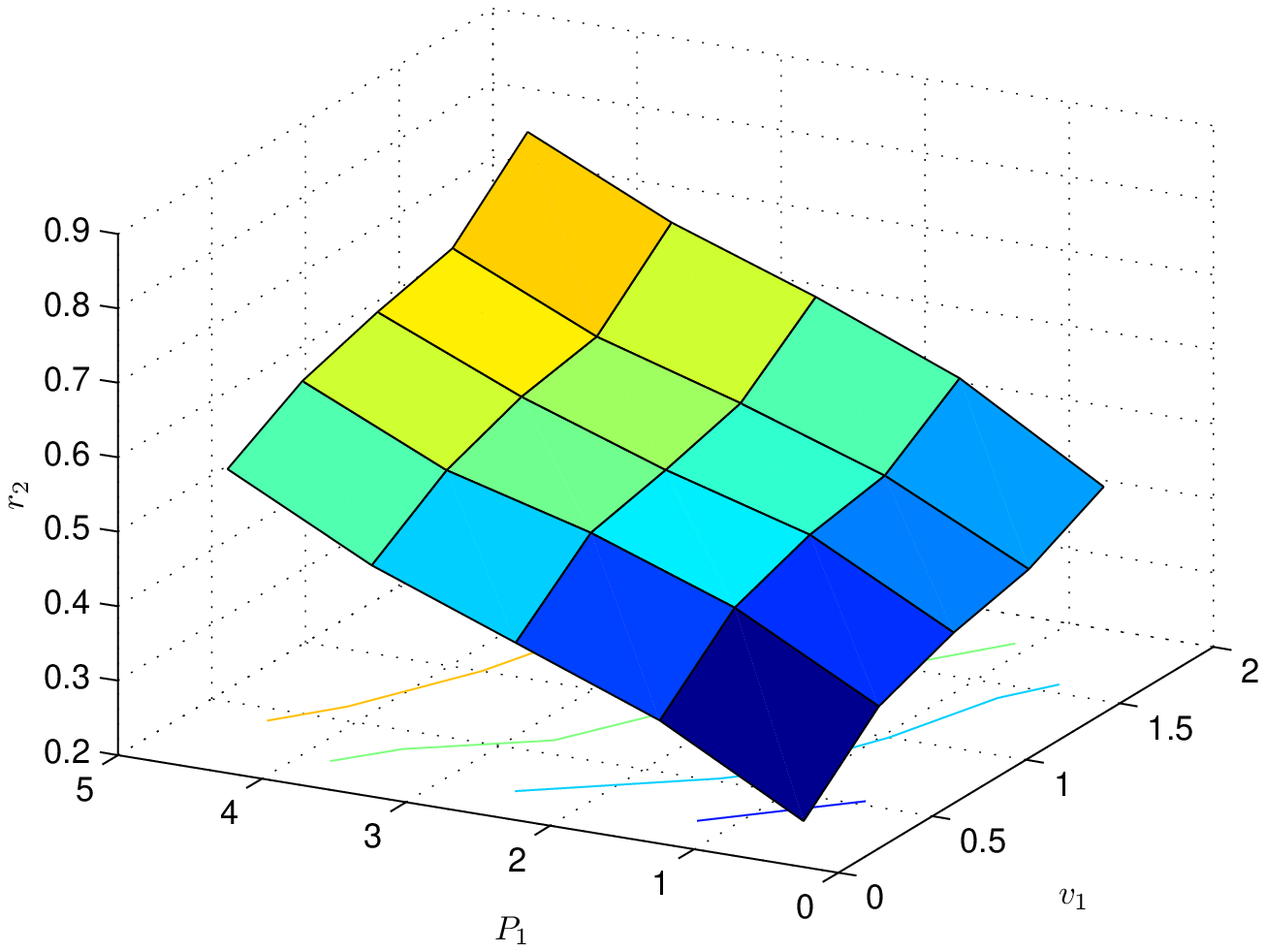} %for one column
\end{center}
\vspace{-10mm}
\caption {The ratio of power allocated for jamming the first transceiver pair over the total jamming power in the optimal jamming strategy versus $P_1$ and $v_1$.} \label{f:NCjamMUltiORT4002}
\vspace{-2mm}
\end{figure}

\emph{Example~3: Jamming TDM based multiple legitimate signals.} A system with two legitimate transceiver pairs and one jammer is considered. It is assumed that the legitimate transmissions are based on TDM. The time division factors $\beta_1$ and $\beta_2$ are both equal to 0.5. The number of antennas at the transmitter and receiver of the first transceiver pair are both 4 while the number of antennas at the transmitter and receiver of the second transceiver pair are both 3. The number of antennas at the jammer is $4$. The power limit of the jammer is $4$ and the noise variances $\sigma_i^2, \forall i$ are equal to 1. We fix the total transmit power of the two legitimate transmitters while changing their individual transmit power to demonstrate the effect of legitimate transmitter power on the rate under jamming. The total transmit power of the two transceiver pairs is fixed at 5 while the transmit power for the first transceiver pair is denoted as $P_1$ $(0<P_1<5)$. The elements of the channels $\mathbf{H}_{i}, \forall i$ are generated from complex Gaussian distribution with zero mean and unit variance. To demonstrate the effect of the quality of the jamming channels, the elements of the jamming channels $\mathbf{H}_{\mathrm{z}1}$ and $\mathbf{H}_{\mathrm{z}2}$ are generated with zero mean and variances $v_1$ $(0<v_1<2)$ and $2-v_1$, respectively. For each combination of $P_1$ and $v_1$,  we use 400 channel realizations and calculate the average $R_\mathrm{ot}^\mathrm{J}$ (obtained using the numerical method in Section~\ref{ss:MultOrtho}) and the average sum-rate without jamming, denoted as $R_\mathrm{ot}^{0}$. Then two ratios are obtained. The first ratio $r_1=1-R_\mathrm{ot}^\mathrm{J}/R_\mathrm{ot}^{0}$ represents the effect of jamming in terms of the decrease of sum-rate in percentage. The second ratio $r_2$ is the ratio of the power allocated for jamming the first target signal over the total jamming power in the optimal jamming strategy.

Figs.~\ref{f:NCjamMUltiORT4001}~and~\ref{f:NCjamMUltiORT4002} demonstrate $r_1$ versus $P_1$ and $v_1$ and $r_2$ versus $v_1$ and $P_1$, respectively. Fig.~\ref{f:NCjamMUltiORT4001} intuitively shows that jamming is more effective, in terms of the percentage of sum-rate reduction, when the jamming channel to the receiver of the transceiver pair with larger transmission power is stronger. Otherwise, the jammer needs to spend a significant amount of jamming power on the transceiver with larger transmission power (since the corresponding jamming channel is weak) in order to minimize the sum-rate. This fact can be seen from Fig.~\ref{f:NCjamMUltiORT4002}. Comparing the above two figures, it can also be seen that jamming is generally more effective, in terms of the percentage of sum-rate reduction, when $P_1$ and $v_1$ are set such that the power allocated for jamming the two targets are about the same.

\section{Conclusion}\label{s:conclu}
\vspace{-1mm}
The general closed-form expression for the optimal solution to the problem of jamming a legitimate communication on a MIMO Gaussian channel is found under the condition that the expression is PSD. The effect of jamming power and jamming channel on the optimal jamming strategy is analyzed. For the case that the PSD condition is not satisfied, a suboptimal solution in closed-form is obtained as an approximation of the optimal solution while a numerical solution is also proposed. It is further shown that the numerical solution, and possibly the closed-form optimal/suboptimal solutions too, can be extended to different scenarios of multi-target jamming after proper modifications. Simulation results for single target jamming demonstrate that the proposed suboptimal solution is very close to the optimal one. For multi-target jamming, the achievable minimum rate under jamming and the jammer's power allocation strategy are illustrated versus the target signal strength and the jamming channel quality.

\section{Appendix}\label{s:appen}

\subsection{Proof of Lemma~1}\label{App:C5prfl1}
%==============================================================================
It is well-known that the function $\log{|\mathbf{I}+\mathbf{A}\mathbf{X}^{-1}|}$ is convex with respect to $\mathbf{X}$ given that $\mathbf{A}$ is PSD \cite{Worstnoise}. Moreover, strong convexity holds if $\mathbf{A}\succ0$. Therefore, the optimal solution can be characterized using the Karush-Kuhn-Tucker (KKT) conditions\cite{ConOpt}.

The Lagrangian for the problem \eqref{e:BJobj} can be written as
\begin{equation}\label{e:BJLagr}
L(\mathbf{X}, \lambda, \mathbf{Z})=\log{|\mathbf{A}+\mathbf{X}|}-\log|\mathbf{X}|+ \lambda (\text{Tr}\{\mathbf{X}\}-1) + \text{Tr}\{\mathbf{X}\mathbf{Z}\}
\end{equation}
where $\lambda$ and $\mathbf{Z}$ are the Lagrange multipliers associated with \eqref{e:BJcons1} and \eqref{e:BJcons2}, respectively. The KKT optimality conditions for the problem \eqref{e:BJ} are then given as
\begin{eqnarray}\label{e:BJKKT}
& \text{Tr}\{\mathbf{X}\}\leq 1,\;
 \mathbf{X} \succeq 0 ,\;
 \lambda\geq 0 \label{e:BJKKT1}\\
& \mathbf{Z}\succeq 0,\;
 \lambda(\text{Tr}\{\mathbf{X}\}- 1)=0, \;
 \text{Tr}\{\mathbf{X}\mathbf{Z}\}=0  \label{e:BJKKT2}\\
& (\mathbf{X}+\mathbf{A})^{\rm -T}- \mathbf{X}^{\rm -T} + \lambda \mathbf{I} + \mathbf{Z}^{\rm T}=\mathbf{0} \label{e:BJKKT3}
\end{eqnarray}
where $(\cdot)^\mathrm{T}$ denotes transpose and $\mathbf{0}$ denotes an all-zero matrix of an appropriate size. It is not difficult to see that $\mathbf{X} \succ 0$ and $\text{Tr}\{\mathbf{X}\}=1$ at optimality. Given that $\mathbf{X} \succ 0$ and $\mathbf{Z}\succeq 0$ at optimality, the condition $\text{Tr}\{\mathbf{X}\mathbf{Z}\}=0$ indicates that $\mathbf{Z}=\mathbf{0}$. Then \eqref{e:BJKKT3} becomes
\begin{equation}\label{e:BJKKT3Eqv1} % 1st equivalent form of {e:BJKKT3}
(\mathbf{X}+\mathbf{A})^{\rm -T}= \mathbf{X}^{\rm -T} - \lambda \mathbf{I}
\end{equation}
which further indicates that
\begin{equation}\label{e:BJKKT3Eqv1} % 1st equivalent form of {e:BJKKT3}
\mathbf{X}+\mathbf{A}= (\mathbf{X}^{\rm -1} - \lambda \mathbf{I})^{-1}.
\end{equation}
Using the matrix inversion lemma \cite{MatrixInvLemma}, the right-hand side of \eqref{e:BJKKT3Eqv1} is equivalent to
\begin{equation}
\mathbf{X}+\mathbf{X}(\mathbf{I}- \lambda \mathbf{X})^{-1} \lambda\mathbf{X}.
\end{equation}
Then \eqref{e:BJKKT3Eqv1} can be written as
\begin{equation}\label{e:BJKKT3Eqv2} % 1st equivalent form of {e:BJKKT3}
\mathbf{A}= \mathbf{X}(\lambda^{-1}\mathbf{I}- \mathbf{X})^{-1} \mathbf{X}.
\end{equation}
Denoting the EVD of $\mathbf{X}$ as $\mathbf{X}=\mathbf{U}_{\rm X}\mathbf{\Lambda}_{\rm X}\mathbf{U}_{\rm X}^{\rm H}$, the expression \eqref{e:BJKKT3Eqv2} can be rewritten as
\begin{equation}\label{e:BJKKT3Eqv3} % 1st equivalent form of {e:BJKKT3}
\mathbf{U}_{\rm X}^{\rm H}\mathbf{A}\mathbf{U}_{\rm X}= \mathbf{\Lambda}_{\rm X}(\lambda^{-1}\mathbf{I}-\mathbf{\Lambda}_{\rm X})^{-1}\mathbf{\Lambda}_{\rm X}.
\end{equation}
Defining $\mathbf{\Lambda}_1\triangleq\mathbf{U}_{\rm X}^{\rm H}\mathbf{A}\mathbf{U}_{\rm X}$, and using the fact that $\mathbf{U}_{\rm X}^{\rm H}\mathbf{A}\mathbf{U}_{\rm X}$ and $\mathbf{A}$ share the same eigenvalues, it can be found that $\mathbf{\Lambda}_1$ contains the eigenvalues of $\mathbf{A}$. Since $\mathbf{U}_{\rm X}^{\rm H}\mathbf{A}\mathbf{U}_{\rm X}$ gives the matrix of eigenvalues of $\mathbf{A}$, it must hold that $\mathbf{U}_{\rm X}=\mathbf{U}_{\rm A}$. Therefore, using $\mathbf{U}_{\rm X}=\mathbf{U}_{\rm A}$, we obtain that
\begin{equation}\label{e:BJKKT3Eqv4} % 1st equivalent form of {e:BJKKT3}
\mathbf{\Lambda}_{\rm A}= \mathbf{\Lambda}_{\rm X}(\lambda^{-1}\mathbf{I}-\mathbf{\Lambda}_{\rm X})^{-1}\mathbf{\Lambda}_{\rm X}
\end{equation}
which gives (recall that $\mathbf{A} \succ 0$ and $\mathbf{X} \succ 0$ at optimality)
\begin{equation}\label{e:BJKKT3Eqv4} % 1st equivalent form of {e:BJKKT3}
\mathbf{\Lambda}_{\rm X}\mathbf{\Lambda}_{\rm A}^{-1}\mathbf{\Lambda}_{\rm X}= \lambda^{-1}\mathbf{I}-\mathbf{\Lambda}_{\rm X}.
\end{equation}
Finally, the following equation
\begin{equation}\label{e:BJKKT3Eqv4} % 1st equivalent form of {e:BJKKT3}
\mathbf{\Lambda}_{\rm X}^2+\mathbf{\Lambda}_{\rm \mathbf{A}}\mathbf{\Lambda}_{\rm \mathbf{X}}= \lambda^{-1}\mathbf{\Lambda}_{\rm A}
\end{equation}
holds, which leads to \eqref{e:BJsolu}.

%==============================================================================
\subsection{Proof of Lemma~2}\label{App:C5prfl2}
%==============================================================================
If $\mathbf{B}$ is PD, the following matrix
\begin{equation}
\bar{\mathbf{B}}=\mathbf{B}+    \kbordermatrix{
~   & r_\mathrm{z} & n_\mathrm{z}-r_\mathrm{z} \cr
    r_\mathrm{z}   & {\mathbf{0}}   &  {\mathbf{0}}  \cr
    n_\mathrm{z}-r_\mathrm{z} &  {\mathbf{0}}  &  {\sigma^2\mathbf{I}} \cr}.
\end{equation}
and its inverse $\bar{\mathbf{B}}^{-1}$ are also PD. Given that $\bar{\mathbf{B}}$ is PD, it can be seen that the two blocks on the diagonal of $\bar{\mathbf{B}}$ are both PD. Then, using block matrix inversion \cite{BLKMatrixInv}, it follows that the first block of $\bar{\mathbf{B}}^{-1}$ is $({\mathbf{B}_{11}}- {\mathbf{B}_{12}}(\sigma^2\mathbf{I}+{\mathbf{B}_{22}})^{-1}{\mathbf{B}_{21}})^{-1}$, which is the inverse of $\mathbf{\tilde{B}}$. Given that $\bar{\mathbf{B}}^{-1}$ is PD, the first block of $\bar{\mathbf{B}}^{-1}$, i.e., the inverse of $\mathbf{\tilde{B}}$ must also be PD. Therefore, $\mathbf{\tilde{B}}$ is also PD. This proves Lemma~2.

%==============================================================================
\subsection{Proof of Theorem~1} \label{App:C5prfT1}
%==============================================================================

Using the definitions \eqref{e:Bblocks}, \eqref{e:tildeB}, \eqref{e:Qhatblockt}, and \eqref{e:OmgzTt}, the objective function in \eqref{e:JOneobjEqv1t} can be rewritten as
\begin{eqnarray}\label{e:JOneobjEqv2}
R^{\rm J}&\hspace{-0.6cm}=&\hspace{-0.4cm}\log{|\mathbf{I}+\mathbf{B}(\mathbf{\tilde{\Omega}}_\mathrm{z}
\mathbf{\tilde{Q}}_\mathrm{z}\mathbf{\tilde{\Omega}}_\mathrm{z}^\mathrm{H}+\sigma^2\mathbf{I})^{-1}|} \nonumber\\
&\hspace{-0.6cm}=&\hspace{-0.4cm}\log{\Bigg|\mathbf{I}+\mathbf{\tilde{\Omega}}_\mathrm{z}^{-1}\mathbf{B}
\mathbf{\tilde{\Omega}}_\mathrm{z}^{-\mathrm{H}}(\mathbf{\tilde{Q}}_\mathrm{z}+\sigma^2\mathbf{\tilde{\Omega}}_\mathrm{z}^{-1}
\mathbf{\tilde{\Omega}}_\mathrm{z}^{-\mathrm{H}})^{-1}\Bigg|}
\nonumber \\
&\hspace{-0.6cm}=&\hspace{-0.4cm}\log \left|\mathbf{I}\! + \! \left[\!\! {\begin{array}{*{20}c}
   {{\mathbf{\Omega}_\mathrm{z}^{+}}} & {\mathbf{0}}\\
   {\mathbf{0}}&{\mathbf{I}}\\
   \end{array} } \!\!\right]^{\!-1}\!\!
   \left[\!\! {\begin{array}{*{20}c}
   {{\mathbf{B}_{11}}\!\!} & {{\mathbf{B}_{12}}\!}\\
   {{\mathbf{B}_{21}}\!\!}&{{\mathbf{B}_{22}}\!}\\
   \end{array} } \!\!\right]\!\!
   \left[\!\! {\begin{array}{*{20}c}
   {{\mathbf{\Omega}_\mathrm{z}^{+}}^\mathrm{H}\!\!} & {\mathbf{0}}\\
   {\mathbf{0}\!\!}&{\mathbf{I}}\\
   \end{array} } \!\!\right]^{\!-1}\!\!\cdot\right. \nonumber \\
&&\hspace{8mm} \left.\bigg(\!\left[\!\! {\begin{array}{*{20}c}
   {\mathbf{Q}^{\prime}_\mathrm{z}} & {\mathbf{0}}\\
   {\mathbf{0}}&{\mathbf{0}}\\
   \end{array} } \!\!\right]
   \!\!+ \sigma^2\!\left[\!\! {\begin{array}{*{20}c}
   {{\mathbf{\Omega}_\mathrm{z}^{+}}^{-1}{\mathbf{\Omega}_\mathrm{z}^{+}}^{-\mathrm{H}}\!\!} & {\mathbf{0}}\\
   {\mathbf{0}\!\!}&{\mathbf{I}}\\
   \end{array} } \!\!\right]\!\bigg)^{\!\!\!-1} \right| \nonumber \\
&\hspace{-0.6cm}=&\hspace{-0.4cm}\log\left|\mathbf{I}+ \left[\!\! {\begin{array}{*{20}c}
   {{{\mathbf{\Omega}_\mathrm{z}^{+}}}^{-1}{\mathbf{B}_{11}}{\mathbf{\Omega}_\mathrm{z}^{+}}}^{-\mathrm{H}\!\!} & {{{\mathbf{\Omega}_\mathrm{z}^{+}}}^{-1}{\mathbf{B}_{12}}}\\
   {{\mathbf{B}_{21}}{{\mathbf{\Omega}_\mathrm{z}^{+}}}^{-\mathrm{H}}\!\!}&{{\mathbf{B}_{22}}}\\
   \end{array} } \!\!\right]\!\cdot\right. \nonumber \\
&&\hspace{12mm}  \left.\left[\!\! {\begin{array}{*{20}c}
   {(\mathbf{Q}^{\prime}_\mathrm{z}+
   \sigma^2{\mathbf{\Omega}_\mathrm{z}^{+}}^{-1}{\mathbf{\Omega}_\mathrm{z}^{+}}^{-\mathrm{H}})^{-1}\!\!} & {\mathbf{0}}\\
   {\mathbf{0}\!\!}&{\frac{1}{\sigma^2}\mathbf{I}}\\
   \end{array} } \!\!\right] \right| \nonumber\\
&\hspace{-0.55cm}=&\hspace{-0.4cm}\log{\left|\left[ \!\!\!{\begin{array}{*{20}c}
   {\mathbf{I}\!+\!{{\mathbf{\Omega}_\mathrm{z}^{+}}}^{-1}{\mathbf{B}_{11}}{{\mathbf{\Omega}_\mathrm{z}^{+}}}^{-\mathrm{H}}
    \mathbf{J}^{-1}\!\!} & {\!\frac{1}{\sigma^2}{{\mathbf{\Omega}_\mathrm{z}^{+}}}^{-1}{\mathbf{B}_{12}}}\\
   {{\mathbf{B}_{21}}{{\mathbf{\Omega}_z^{+}}}^{-\mathrm{H}}\mathbf{J}^{-1}\!\!} & {\!\mathbf{I}\!+\!\frac{1}{\sigma^2}{\mathbf{B}_{22}}}\\
   \end{array} } \!\!\!\right]\right|}
\end{eqnarray}
where in the last step $\mathbf{J}\triangleq\mathbf{Q}^{\prime}_\mathrm{z}+\sigma^2{\mathbf{\Omega}_\mathrm{z}^{+}}^{-1}{\mathbf{\Omega}_\mathrm{z}^{+}}^{-\mathrm{H}}$.

Since the matrix $\mathbf{H}_\mathrm{r}\mathbf{Q}_\mathrm{s}\mathbf{H}_\mathrm{r}^\mathrm{H}$ is PD, $\mathbf{B}$, and consequently $\mathbf{B}_{11}$ and $\mathbf{B}_{22}$ in \eqref{e:Bblocks}, are all PD. The rate $R^{\rm J}$ in \eqref{e:JOneobjEqv2} can be simplified as
\begin{equation}\label{e:Rsplit}
R^{\rm J}=R^0+\bar{R}^\mathrm{J}
\end{equation}
where
\begin{equation}\label{e:R0}
R^0=\log{\bigg|\mathbf{I}+\frac{1}{\sigma^2}{\mathbf{B}_{22}}\bigg|}
\end{equation}
is the part of rate that is not affected by jamming which is non-zero if $r_\mathrm{z}<n_\mathrm{r}$ and
\begin{eqnarray}\label{e:RJ}
&&\hspace{-10mm}\bar{R}^\mathrm{J}=\mathrm{log}\bigg|\mathbf{I}+{{\mathbf{\Omega}_\mathrm{z}^{+}}}^{-1}{\mathbf{B}_{11}}
{\mathbf{\Omega}_\mathrm{z}^{+}}^{-\mathrm{H}}\mathbf{J}^{-1} \nonumber\\
&&\hspace{0mm}-\frac{1}{\sigma^2}{{\mathbf{\Omega}_\mathrm{z}^{+}}}^{-1}{\mathbf{B}_{12}}(\mathbf{I}+\frac{1}{\sigma^2}{\mathbf{B}_{22}})^{-1}
{\mathbf{B}_{21}}{{\mathbf{\Omega}_\mathrm{z}^{+}}}^{-\mathrm{H}}\mathbf{J}^{-1} \bigg|
\end{eqnarray}
is the part of the rate that is affected by jamming. Therefore, the minimization of $R^{\rm J}$ in \eqref{e:JOneobj} is equivalent to minimizing  $\bar{R}^\mathrm{J}$. Using the definition of $\mathbf{\tilde{B}}$ in \eqref{e:tildeB}, $\bar{R}^\mathrm{J}$ can be rewritten as
\begin{equation}\label{e:RJEqv1}
\bar{R}^\mathrm{J}=\log{|\mathbf{I}+{{\mathbf{\Omega}_\mathrm{z}^{+}}}^{-1}{\mathbf{\tilde{B}}}
{\mathbf{\Omega}_\mathrm{z}^{+}}^{-\mathrm{H}}(\mathbf{Q}^{\prime}_\mathrm{z}+\sigma^2{\mathbf{\Omega}_\mathrm{z}^{+}}^{-1}
{\mathbf{\Omega}_\mathrm{z}^{+}}^{-\mathrm{H}})^{-1}|}.
\end{equation}
Using Lemma~2, it can be seen that $\mathbf{\tilde{B}}$ is PD when $\mathbf{B}$ is PD. Then, Lemma~1 can be used to find such $\mathbf{Q}^{\prime}$ that minimizes \eqref{e:RJEqv1} subject to the trace constraint $\text{Tr}\{\mathbf{Q}^{\prime}_\mathrm{z}\}\leq P_\mathrm{z}$. 
Using \eqref{e:BJsolu}, the definition $\mathbf{\tilde{A}}\triangleq{\mathbf{\Omega}_\mathrm{z}^{+}}^{\!-1}\mathbf{\tilde{B}} {{\mathbf{\Omega}_\mathrm{z}^{+}}}^{\!-\mathrm{H}}$, and the EVD $\mathbf{\tilde{A}}=\mathbf{U}_{\mathbf{\tilde{A}}}\mathbf{\Lambda}_{\mathbf{\tilde{A}}}\mathbf{U}_{\mathbf{\tilde{A}}}^\mathrm{H}$, the matrix $\mathbf{Q}^{\prime}_\mathrm{z}$ that minimizes \eqref{e:RJEqv1}, or equivalently \eqref{e:RJ}, subject to  $\text{Tr}\{\mathbf{Q}^{\prime}_\mathrm{z}\}\leq P_\mathrm{z}$ can be found as
\begin{equation}\label{e:QprimeOpt1}
\mathbf{Q}^{\prime}_\mathrm{z}\!=\!\mathbf{U}_{\mathbf{\tilde{A}}}\sqrt{\frac{1}{\lambda}\mathbf{\Lambda}_{\mathbf{\tilde{A}}}\!+\!
\frac{1}{4}\mathbf{\Lambda}_{\mathbf{\tilde{A}}}^2}\mathbf{U}_{\mathbf{\tilde{A}}}^{\rm H}\!-\!{\mathbf{\Omega}_\mathrm{z}^{\!+}}^{\!-1}\bigg(\frac{1}{2}\mathbf{\tilde{B}}\!+\! \sigma^2\mathbf{I}\bigg)
{\mathbf{\Omega}_\mathrm{z}^{\!+}}^{\!-\mathrm{H}}
\end{equation}
under the condition that the above $\mathbf{Q}^{\prime}_\mathrm{z}$ is PSD. Here $\lambda$ is chosen such that $\text{Tr}\{\mathbf{Q}^{\prime}_\mathrm{z}\}=P_\mathrm{z}$. 

%=============================================================================
\subsection{Proof of Theorem~2} \label{App:C5prfT2}
%=============================================================================
The proof follows the same route as the proof of Theorem~1 in Subsection~\ref{App:C5prfT1} till the expression \eqref{e:RJEqv1}. Then, using \eqref{e:AEVDblock}, the $\bar{R}^{\rm J}$ in \eqref{e:RJEqv1} can be rewritten as
\begin{eqnarray}\label{e:RJEqv2}
\bar{R}^{\rm J}\!\!\!\!&=&\!\!\!\!\log{|\mathbf{I}+ \mathbf{\tilde{A}} (\mathbf{Q}^{\prime}_\mathrm{z}+\sigma^2{\mathbf{\Omega}_\mathrm{z}^{+}}^{-1}{\mathbf{\Omega}_\mathrm{z}^{+}}^\mathrm{-H})^{-1}|} \nonumber\\
\!\!\!\!&=&\!\!\!\! \log{\left|\mathbf{I}+ \left[ {\begin{array}{*{20}c}
   {\!\!\mathbf{U}_{\mathbf{\tilde{A}}1}\!\!} & {\!\!\mathbf{U}_{\mathbf{\tilde{A}}2}\!\!}\\
   \end{array} } \right]
   \left[ {\begin{array}{*{20}c}
   {\!\mathbf{\Lambda}_\mathbf{\tilde{A}}^{+}\!} & {\!\mathbf{0}\!}\\
   {\!\mathbf{0}\!}&{\!\mathbf{0}\!}\\
   \end{array} } \right]
   \left[ {\begin{array}{*{20}c}
   {\!\mathbf{U}_{\mathbf{\tilde{A}}1}^\mathrm{H}}\! \\
   {\!\mathbf{U}_{\mathbf{\tilde{A}}2}^\mathrm{H}}\! \\
   \end{array} } \right]
   {\mathbf{Q}_\mathrm{z}^{\prime\prime}}^{-1}\right|}   \nonumber\\
\!\!\!\!&=&\!\!\!\!\!\log\left|\mathbf{I}+
   \left[ {\begin{array}{*{20}c}
   {\!\mathbf{\Lambda}_\mathbf{\tilde{A}}^{+}\!} & {\!\mathbf{0}\!}\\
   {\!\mathbf{0}\!}&{\!\mathbf{0}\!}\\
   \end{array} } \right]
   \bigg(\left[ {\begin{array}{*{20}c}
   {\mathbf{U}_{\mathbf{\tilde{A}}1}^\mathrm{H}} \\
   {\mathbf{U}_{\mathbf{\tilde{A}}2}^\mathrm{H}} \\
   \end{array} } \right]
   \mathbf{Q}_\mathrm{z}^{\prime\prime}
   \left[ {\begin{array}{*{20}c}
   {\!\!\mathbf{U}_{\mathbf{\tilde{A}}1}\!\!} & {\!\!\mathbf{U}_{\mathbf{\tilde{A}}2}\!\!}\\
   \end{array} } \right]  \bigg)^{\!\!-1}\right|  \nonumber \\
\!\!\!\!&=&\!\!\!\!\!\log\left|\mathbf{I}+
     \left[ {\begin{array}{*{20}c}
   {\!\mathbf{\Lambda}_\mathbf{\tilde{A}}^{+}\!} & {\!\mathbf{0}\!}\\
   {\!\mathbf{0}\!}&{\!\mathbf{0}\!}\\
   \end{array} } \right]
   \left[ {\begin{array}{*{20}c}
   {\!\mathbf{U}_{\mathbf{\tilde{A}}1}^\mathrm{H}\mathbf{Q}_\mathrm{z}^{\prime\prime}\mathbf{U}_{\mathbf{\tilde{A}}1}\!} & {\!\mathbf{U}_{\mathbf{\tilde{A}}1}^\mathrm{H}\mathbf{Q}_\mathrm{z}^{\prime\prime}\mathbf{U}_{\mathbf{\tilde{A}}2}\!}\\
   {\!\mathbf{U}_{\mathbf{\tilde{A}}2}^\mathrm{H}\mathbf{Q}_\mathrm{z}^{\prime\prime}\mathbf{U}_{\mathbf{\tilde{A}}1}\!} & {\!\mathbf{U}_{\mathbf{\tilde{A}}2}^\mathrm{H}\mathbf{Q}_\mathrm{z}^{\prime\prime}\mathbf{U}_{\mathbf{\tilde{A}}2}\!}   \\
   \end{array} } \right]^{\!-1}\!\right|   \nonumber \\
\!\!\!\!&=&\!\!\!\!\log{\left|\mathbf{I}+
   \left[ {\begin{array}{*{20}c}
   {\mathbf{\Lambda}_\mathbf{\tilde{A}}^{+}} & {\mathbf{0}}\\
   {\mathbf{0}}&{\mathbf{0}}\\
   \end{array} } \right]
   \left[ {\begin{array}{*{20}c}
   {\mathbf{F}_1^{-1}} & {\mathbf{F}_{12}}\\
   {\mathbf{F}_{21}}&{\mathbf{F}_2^{-1}}\\
   \end{array} } \right]
   \right|}\nonumber \\
\!\!\!\!&=&\!\!\!\! \log{\bigg|\mathbf{I}+ \mathbf{\Lambda}_\mathbf{\tilde{A}}^{+} \mathbf{F}_1^{-1}\bigg|}
\end{eqnarray}
where $\mathbf{Q}_\mathrm{z}^{\prime\prime}\triangleq \mathbf{Q}^{\prime}_\mathrm{z}+\sigma^2{\mathbf{\Omega}_\mathrm{z}^{+}}^{-1}{\mathbf{\Omega}_\mathrm{z}^{+}}^\mathrm{-H}$ in the second step. The result on block matrix inversion \cite{BLKMatrixInv} is used in the last step, in which
\begin{equation}\label{e:F1}
\mathbf{F}_1\triangleq \mathbf{F}_1^1-\mathbf{F}_1^2
\end{equation}
with $\mathbf{F}_1^1$ and $\mathbf{F}_1^2$ given by
\begin{eqnarray}
&&\hspace{-10mm}\mathbf{F}_1^1\triangleq\mathbf{U}_{\mathbf{\tilde{A}}1}^\mathrm{H}\mathbf{Q}_\mathrm{z}^{\prime\prime}
\mathbf{U}_{\mathbf{\tilde{A}}1}\\ &&\hspace{-10mm}\mathbf{F}_1^2\triangleq\mathbf{U}_{\mathbf{\tilde{A}}1}^\mathrm{H}\mathbf{Q}_\mathrm{z}^{\prime\prime}
\mathbf{U}_{\mathbf{\tilde{A}}2}
(\mathbf{U}_{\mathbf{\tilde{A}}2}^\mathrm{H}\mathbf{Q}_\mathrm{z}^{\prime\prime}\mathbf{U}_{\mathbf{\tilde{A}}2}
)^{-1}\mathbf{U}_{\mathbf{\tilde{A}}2}^\mathrm{H}\mathbf{Q}_\mathrm{z}^{\prime\prime}
\mathbf{U}_{\mathbf{\tilde{A}}1}
\end{eqnarray}
and
\begin{eqnarray}
&&\hspace{-5mm}\mathbf{F}_{12}\triangleq-(\mathbf{U}_{\mathbf{\tilde{A}}1}^\mathrm{H}\mathbf{Q}_\mathrm{z}^{\prime\prime}\mathbf{U}_{\mathbf{\tilde{A}}1} )^{-1}\mathbf{U}_{\mathbf{\tilde{A}}1}^\mathrm{H}\mathbf{Q}_\mathrm{z}^{\prime\prime}
\mathbf{U}_{\mathbf{\tilde{A}}2}\mathbf{F}_2^{-1}\\
&&\hspace{-5mm}\mathbf{F}_{21}\triangleq-(\mathbf{U}_{\mathbf{\tilde{A}}2}^\mathrm{H}\mathbf{Q}_\mathrm{z}^{\prime\prime}\mathbf{U}_{\mathbf{\tilde{A}}2}
)^{-1}\mathbf{U}_{\mathbf{\tilde{A}}2}^\mathrm{H}\mathbf{Q}_\mathrm{z}^{\prime\prime}
\mathbf{U}_{\mathbf{\tilde{A}}1}\mathbf{F}_1^{-1}\\
&&\hspace{-5mm}\mathbf{F}_2\triangleq \mathbf{U}_{\mathbf{\tilde{A}}2}^\mathrm{H}\mathbf{Q}_\mathrm{z}^{\prime\prime}\mathbf{U}_{\mathbf{\tilde{A}}2} \nonumber \\
&&\hspace{3mm}-\mathbf{U}_{\mathbf{\tilde{A}}2}^\mathrm{H}\mathbf{Q}_\mathrm{z}^{\prime\prime}\mathbf{U}_{\mathbf{\tilde{A}}1} (\mathbf{U}_{\mathbf{\tilde{A}}1}^\mathrm{H}\mathbf{Q}_\mathrm{z}^{\prime\prime}\mathbf{U}_{\mathbf{\tilde{A}}1} )^{\!-1}\mathbf{U}_{\mathbf{\tilde{A}}1}^\mathrm{H}\mathbf{Q}_\mathrm{z}^{\prime\prime}
\mathbf{U}_{\mathbf{\tilde{A}}2}.
\end{eqnarray}

Recalling the optimization problem \eqref{e:BJ}, it can be seen from the last step of \eqref{e:RJEqv2} that $\bar{R}^{\rm J}$ is not minimized if the trace of $\mathbf{F}_1$ can be increased under the jammer's power constraint. Therefore, a necessary condition for minimizing \eqref{e:RJEqv2} is that the trace of $\mathbf{F}_1$ is maximized given the trace constraint of $\mathbf{Q}_\mathrm{z}^{\prime}$.

Considering the fact that $\text{Tr}\{\mathbf{U}_{\mathbf{\tilde{A}}1}^\mathrm{H}\mathbf{Q}^{\prime\prime}_\mathrm{z}\mathbf{U}_{\mathbf{\tilde{A}}1}\}
\leq \text{Tr}\{\mathbf{Q}^{\prime\prime}_\mathrm{z}\}$ and that $\mathbf{F}_1^2$ is PSD, maximizing $\text{Tr}\{\mathbf{F}_1\}$ requires that $\mathbf{Q}_\mathrm{z}^{\prime\prime}$ must have the following form
\begin{equation}\label{QprimeOpt2}
\mathbf{Q}_\mathrm{z}^{\prime\prime}= \mathbf{U}_{\mathbf{\tilde{A}}1}\mathbf{D}_\mathrm{x}\mathbf{U}_{\mathbf{\tilde{A}}1}^\mathrm{H}
\end{equation}
in which $\mathbf{D}_\mathrm{x}$ is an $r_\mathbf{\tilde{A}}\times r_\mathbf{\tilde{A}}$ PSD matrix to be determined. The matrix $\mathbf{D}_\mathrm{x}$ should satisfy the constraint $\text{Tr}\{\mathbf{D}_\mathrm{x}\}\leq P_\mathrm{z}+ \sigma^2\text{Tr}\{{\mathbf{\Omega}_\mathrm{z}^{+}}^{-1}{\mathbf{\Omega}_\mathrm{z}^{+}}^\mathrm{-H}\}$.

Using \eqref{QprimeOpt2}, $\mathbf{F}_1^2$ is equal to $\mathbf{0}$ and $\mathbf{F}_1$ in \eqref{e:F1} is equal to $\mathbf{D}_\mathrm{x}^{-1}$.
Consequently, \eqref{e:RJEqv2} can be rewritten as
\begin{eqnarray}\label{e:RDx}
R^{\rm J}=\log{|\mathbf{I}+ \mathbf{\Lambda}_\mathbf{\tilde{A}}^{+}\mathbf{D}_\mathrm{x}^{-1}|}.
\end{eqnarray}
Therefore, the matrix $\mathbf{Q}_\mathrm{z}^{\prime\prime}$ in \eqref{QprimeOpt2} corresponds to spreading the power (including jamming power and noise power) on the eigen-channels corresponding to the positive eigenvalues of $\mathbf{\tilde{A}}$. Indeed, `spilling' power on the null space of $\mathbf{\tilde{A}}$ cannot be optimal.

Using the result from Lemma~1, the optimal $\mathbf{D}_\mathrm{x}$ is given as
\begin{equation}\label{e:Dx}
\mathbf{D}_\mathrm{x}=\! \sqrt{\frac{1}{\lambda}\mathbf{\Lambda}_\mathbf{\tilde{A}}^{+}\!+\!
\frac{1}{4}{\mathbf{\Lambda}_{\mathbf{\tilde{A}}}^{+}}^2} \!-\! \frac{1}{2}\mathbf{\Lambda}_\mathbf{\tilde{A}}^{+}.
\end{equation}
Accordingly, the optimal $\mathbf{Q}^{\prime}$ is given as
\begin{eqnarray}\label{e:QprimeOpt2}
\mathbf{Q}_\mathrm{z}^{\prime}\!\hspace{-3mm}&=&\hspace{-3mm} \! \!\mathbf{U}_{\mathbf{\tilde{A}}1}\!\sqrt{\!\frac{1}{\lambda}\mathbf{\Lambda}_\mathbf{\tilde{A}}^{+}\!\!+\!\!
\frac{1}{4}{\mathbf{\Lambda}_{\mathbf{\tilde{A}}}^{+}}^2\!}\mathbf{U}_{\mathbf{\tilde{A}}1}^\mathrm{H} \!\nonumber \\
&&-\! \frac{1}{2}\mathbf{U}_{\mathbf{\tilde{A}}1}\!\mathbf{\Lambda}_\mathbf{\tilde{A}}^{+}\!\mathbf{U}_{\mathbf{\tilde{A}}1}^\mathrm{H} \!\!-\!\sigma^2{\mathbf{\Omega}_\mathrm{z}^{+}}^{\!\!-1}\!{\mathbf{\Omega}_\mathrm{z}^{+}}^\mathrm{\!-H}
\end{eqnarray}
if the above $\mathbf{Q}^{\prime}$ is PSD, where $\lambda$ is chosen such that $\text{Tr}\{\mathbf{Q}_z^{\prime}\}= P_\mathrm{z}$.

%==============================================================================
\subsection{Proof of Lemma~3}\label{App:C5prfl3}
%==============================================================================
The four-step procedure in Table~\ref{t:OptbyAlg} uses the sequential parametric convex approximation method \cite{ConvexRela}. The convergence of this method to optimality is proved in \cite{ConvexRela} assuming that the convex relaxations (in our case, the right-hand side of \eqref{e:LPOptsubcons1}) are ``convex upper estimate functions'' of the right-hand side of the original nonconvex constraints (in our case, the right-hand side of \eqref{e:LPOpteqcons1}). Therefore, it is sufficient to prove that \begin{eqnarray}\label{e:Relacon}
&\hspace{-1cm}\log{|\mathbf{Q}^{\prime}_\mathrm{z}+ \mathbf{D}_0+ \mathbf{\tilde{A}}|} \leq  \log{|{\mathbf{Q}^{\prime}}^{\dag}_\mathrm{z}\! +\!\mathbf{D}_0 \!+\! \mathbf{\tilde{A}}|} \!+\!\text{Tr}\{\big({\mathbf{Q}^{\prime}}^{\dag}_\mathrm{z} \!+\! \nonumber\\
&\hspace{8mm}  \mathbf{D}_0\!+\! \mathbf{\tilde{A}}\big)^{\!-1}\mathbf{Q}^{\prime}_\mathrm{z}\}-\! \text{Tr}\{\big({\mathbf{Q}^{\prime}}^{\dag}_\mathrm{z} \!+\! \mathbf{D}_0 \!+\! \mathbf{\tilde{A}}\big)^{\!-1}{\mathbf{Q}^{\prime}}^{\dag}_\mathrm{z}\}
\end{eqnarray}
for all $\mathbf{Q}^{\prime}_\mathrm{z}$ and ${\mathbf{Q}^{\prime}}^{\dag}_\mathrm{z}$ which are PD and satisfy \eqref{e:LPOpteqcons2}, and that the right-hand side of \eqref{e:Relacon} is convex and continuously differentiable with respect to $\mathbf{Q}^{\prime}_\mathrm{z}$ given  ${\mathbf{Q}^{\prime}}^{\dag}_\mathrm{z}$. It is not difficult to see that the latter condition is satisfied. Thus, we only need to prove the first point. Using Taylor expansion, it can be shown that the right-hand side of \eqref{e:Relacon} is the tangent of the function $f(\mathbf{Q}^{\prime}_\mathrm{z})=\log{|\mathbf{Q}^{\prime}_\mathrm{z}+ \mathbf{D}_0+ \mathbf{\tilde{A}}|}$ at $\mathbf{Q}^{\prime}_\mathrm{z}={\mathbf{Q}^{\prime}}^{\dag}_\mathrm{z}$ \cite{Dattorro}. Recalling the fact that the function $f(\mathbf{Q}^{\prime}_\mathrm{z})=\log{|\mathbf{Q}^{\prime}_\mathrm{z}+ \mathbf{D}_0+ \mathbf{\tilde{A}}|}$ is strictly concave when $\mathbf{Q}^{\prime}_\mathrm{z}\succ 0$, it can be seen that \eqref{e:Relacon} is satisfied for all valid $\mathbf{Q}^{\prime}_\mathrm{z}$ and ${\mathbf{Q}^{\prime}}^{\dag}_\mathrm{z}$.

\end{document}